\documentclass[aps, prl, superscriptaddress, citeautoscript, amsmath, amssymb, showkeys, reprint, floatfix]{revtex4-2}

\usepackage{xr}
\makeatletter
\newcommand*{\addFileDependency}[1]{
  \typeout{(#1)}
  \@addtofilelist{#1}
  \IfFileExists{#1}{}{\typeout{No file #1.}}
}
\makeatother

\newcommand*{\myexternaldocument}[1]{
    \externaldocument[S]{#1}
    \addFileDependency{#1.tex}
    \addFileDependency{#1.aux}
}

\myexternaldocument{Supplement} 

\usepackage{graphicx}
\usepackage{dcolumn}
\usepackage{bm}
\usepackage{hyperref}
\usepackage{xcolor}
\usepackage{blindtext}
\usepackage[separate-uncertainty=true]{siunitx}

\begin{document}

\title{Steric Engineering of Exciton Fine Structure in 2D Perovskites}

\author{Mateusz Dyksik}
\email[]{mateusz.dyksik@pwr.edu.pl}
\affiliation{Department of Experimental Physics, Faculty of Fundamental Problems of Technology, Wroclaw University of Science and Technology, Wroclaw, Poland}

\author{Michal Baranowski}
\affiliation{Department of Experimental Physics, Faculty of Fundamental Problems of Technology, Wroclaw University of Science and Technology, Wroclaw, Poland}

\author{Joshua J. P. Thompson}
\affiliation{Department of Materials Science and Metallurgy, University of Cambridge, Cambridge CB3 0FS, United Kingdom}

\author{Zhuo Yang}
\affiliation{The Institute for Solid State Physics, The University of Tokyo, Kashiwanoha 5-1-5, Kashiwa, Chiba, Japan}

\author{Martha Rivera Medina}
\affiliation{Zernike Institute for Advanced Materials, University of Groningen, Nijenborgh 4, 9747 AG Groningen, The Netherlands}


\author{Maria Antonietta Loi}
\affiliation{Zernike Institute for Advanced Materials, University of Groningen, Nijenborgh 4, 9747 AG Groningen, The Netherlands}

\author{Ermin Malic}
\affiliation{Department of Physics, Philipps-Universit\"{a}t Marburg, Renthof 7, 35032 Marburg}

\author{Paulina Plochocka}
\email{paulina.plochocka@lncmi.cnrs.fr}
\affiliation{Department of Experimental Physics, Faculty of Fundamental Problems of Technology, Wroclaw University of Science and Technology, Wroclaw, Poland}
\affiliation{Laboratoire National des Champs Magn\'etiques Intenses, EMFL, CNRS UPR 3228, University Grenoble Alpes, University Toulouse, University Toulouse 3, INSA-T, Grenoble and Toulouse, France}


\date{\today}
        
\begin{abstract}

We provide a comprehensive study of excitonic properties of 2D layered perovskites, with an emphasis on understanding and controlling the exciton fine structure. First, we present an overview of the optical properties, discussing the challenges in determining the band gap and exciton binding energies. Through magneto-optical spectroscopic measurements (up to $B=140$\,T), we establish scaling laws for exciton binding energy as a function of the band gap and the diamagnetic coefficient. Using an in-plane magnetic field we examine the exciton fine structure for various 2D perovskites to measure the energy splitting between the excitonic levels. We correlate the exciton fine structure and exchange interaction with structural parameters, employing an effective mass model, to highlight the role of steric effects on the exchange interaction. Our findings reveal that lattice distortions, introduced by organic spacers, significantly influence the exchange interaction, driving a tunable energy spacing between dark and bright excitons. This unique feature of 2D perovskites, not present in other semiconductors, offers a novel tuning mechanism for exciton control, making these materials highly promising for efficient light emitters and advanced quantum technologies.
\end{abstract}

\keywords{}
                             
\maketitle

\section{Introduction}
Over the last decade, organic-inorganic metal-halide perovskite fever\cite{Buchanan2020fever} triggered the investigation of their different derivatives\cite{Ruddlesden1958compound,Dion1981Nouvelles,Jacobson1985Interlayer,Mitzi1994Conducting,Dyksik2020influence} in an attempt to overcome the environmental stability issue typical for 3D counterparts\cite{tsai2016high,asghar2017device}. Among numerous classes of materials Ruddlesden-Popper two-dimensional (2D) layered perovskites have demonstrated a strong potential due to their enhanced stability under ambient conditions, exhibiting reasonable power conversion efficiency\cite{tsai2016high,Shao2022Over} and demonstrating superior light emission properties\cite{chen20182d, yan2018recent, van2018recent, gong2018electron,Qin2020stable}. In addition, these natural quantum wells bridge, in an unexpected way, the features of organic and inorganic semiconductors forming a \emph{hybrid} system characterized by a complex, optical response\cite{urban2020revealing,straus2016direct,baranowski2019phase,dyksik2020broad,Neutzner2018exciton} driven by the particularly strong excitonic and polaronic effects\cite{straus2018electrons,blancon2018scaling,dyksik2021tuning,Dyksik2024polaron,thouin2019phonon,parra2022large}.

Although the concept of excitons -- elementary excitations in classic semiconductors -- has been known and studied for decades\cite{Yu2010cardona}, understanding and accurately determining the properties of these quasi-particles in 2D perovskites has proven to be a significant challenge\cite{Yaffe2015excitons,Leppert2024excitons}. This is mostly related to the complex optical responses of 2D perovskites, that have resulted in numerous controversies, related to, for instance, the order of the excitonic states within a fine structure\cite{Posmyk2024exciton,Quarti2024exciton, ben2020multiband, steger2022optical, ema2006huge, DeCrescent2020evenparity}, or the nature of the observed multiple absorption and emission peaks\cite{Do2020, Neutzner2018exciton, Dyksik2024polaron, thouin2019phonon, urban2020revealing, canet2022revealing, Posmyk2022quantification, posmyk2023fine, fang2020band}. 
As a consequence, despite intensive studies, we have only a limited understanding of the emission spectra, where the appearance of multiple peaks has been attributed to bi-exciton formation\cite{fang2020band}, in-gap states\cite{Kahmann2021photophysics}, electron-phonon coupling\cite{Moral2020influence,Feldstein2020microscopic}, and magnetic dipole transitions\cite{DeCrescent2020evenparity}. The challenges in quantifying key excitonic properties of 2D perovskites have significant implications for the performance of future devices. For instance, the exciton binding energy determines the balance between the free charge carrier and the exciton populations\cite{Gelvez-Rueda2017interconversion, simbula2022direct}, which in turn impacts the efficiency of charge transport, a key factor in solar cell performance\cite{Stranks2013electron}. This also influences the oscillator strength of optical transitions and the exciton lifetime. Moreover, the understanding and ability to control and modify the exciton fine structure is essential for advanced quantum devices, such as single-photon sources, entangled photon sources and quantum teleportation systems\cite{canneson2017negatively, tamarat2019ground, xu2018long,Stevenson2006semiconductor,BassoBasset2021quantum, cai2023zero, han2022lattice}. Finally, the interplay between different excitonic states and phonons turned out to be crucial in understanding the exceptionally strong emission exhibited by low-dimensional perovskites \cite{Dyksik2021brightening,Thompson2024phononbottleneck,tamarat2019ground,Wang2022thickness}.

The exploration of excitonic properties in 2D perovskites has predominantly relied on case studies, impeding a more generalized picture of excitons across the plethora of available compositions\cite{du2017two}. In this work we attempt to remedy this issue, presenting a broad-scale investigation and description of the excitonic properties of the thinnest representatives of 2D layered perovskites (described by the general formula A$_2$MX$_4$ where A is an organic spacer, M is a metal cation and X is halide anion)\cite{chen20182d}. We investigate a diverse sample set featuring different metal cations (Pb, Sn), halide anions (I, Br), and organic spacers (PEA -- phenylethylammonium, BA -- butylammonium). The choice of metal cations and halide anions tunes the optical response in a broad spectral range from approximately 2\,eV to 3.5\,eV, while the two distinct organic spacers induce significant differences in octahedral distortions. This provides a robust foundation for comprehensively understanding of excitons in the most commonly investigated 2D perovskite compounds. Exploiting magneto-optical spectroscopy we ascertain key parameters of the electronic properties, including band gap energy, exciton binding energy, and the fine structure characteristics of the exciton manifold for a broad range of 2D perovskites with varying organic sublattices alongside the metal and halide components. Our findings reveal a linear relationship between the exciton binding energy and band gap energy, providing a phenomenological scaling law. Furthermore, we determine the exciton fine structure for multiple 2D layered perovskites characterized by a different lattice distortions imposed by organic spacers. Surprisingly, it has a non-trivial impact on the exchange interaction allowing the energy spacing between dark and bright excitons to be tuned. This tuning knob, not available in classic semiconductors, makes 2D perovskites a unique material system where the exciton manifold can be controlled via the steric effect. 

The manuscript is organized as follows; (i) we begin with the overview of the optical response of 2D layered perovskites. We discuss the related challenges in determining band gap and 1s exciton energy. Based on magneto-optical spectroscopic studies we derive scaling laws for exciton binding energy $E_b$ on the band gap energy $E_g$ and the diamagnetic coefficient $c_0$. (ii) Then we introduce the exciton fine structure in 2D layered perovskites. Using the magneto-transmission technique in Voigt geometry we obtain energy splitting between excitonic levels. (iii) in the latter section we correlate the determined exciton fine structures and exchange interaction with structural parameters. We analyse our results within the effective mass models revealing the importance of the steric effect on the exchange interaction.

\section{Exciton binding energy and band gap determination}

\begin{figure*}
    \centering
    \includegraphics[width=1\linewidth]{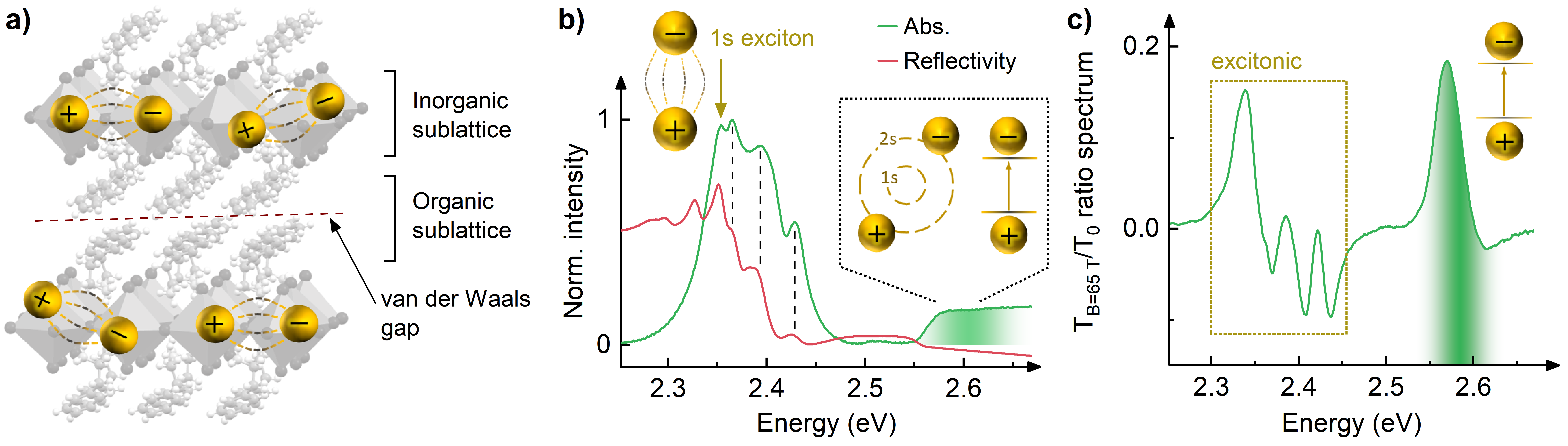}
    \caption{a) Crystal structure of 2D layered perovskite (PEA)$_2$PbI$_4$ with a schematic representation of excitons residing in the inorganic slabs. b) The comparison of absorbance and reflectivity spectra of (PEA)$_2$PbI$_4$. Dashed vertical lines connect the respective optical transition in both spectra. c) Transmission ratio spectrum of (PEA)$_2$PbI$_4$: transmission measured at the high magnetic field of \SI{65}{T} divided by zero-field spectrum.  }
    \label{fig:fig1_abs_R_structure}
\end{figure*}

The structure of typical 2D layered perovskites is presented in \textbf{Figure}\,\ref{fig:fig1_abs_R_structure}a. The crystal structure comprises of organic and inorganic sublattices (hence hybrid perovskite), forming a quantum-well-like arrangement\cite{Mauck2019excitons,Cao2015homologous,Papavassiliou1994structural}. In such a configuration the organic sublattice serves as a barrier for the charge carriers\cite{Blancon2020semiconductor,Even2014understanding}. Additionally, the organic spacer possesses a lower dielectric constant when compared with the well material\cite{katan2019quantum}, introducing a dielectric confinement effect in 2D layered perovskites\cite{katan2019quantum}. As a result of both quantum and dielectric confinements the well material hosts strongly bound excitons confined within the inorganic slab, with binding energies in the range of hundreds of meV\cite{blancon2018scaling,dyksik2021tuning,Yaffe2015excitons}, as schematically depicted in \textbf{Figure}\,\ref{fig:fig1_abs_R_structure}a. 

The significant exciton binding energy $E_b$ separates the excitonic signatures from the quasi-particle band gap $E_g$. Thus, the $E_b$ should, in principle, be straightforward to determine by direct measurements of the 1s exciton energy $E_{1s}$ and the band gap-related step-like features $E_g$. Then $E_b=E_g-E_{1s}$. However, this approach is challenging in the case of 2D perovskites. 

A typical absorption and reflectivity spectrum of a representative 2D perovskite, (PEA)$_2$PbI$_4$, is shown in \textbf{Figure}\,\ref{fig:fig1_abs_R_structure}b (details on sample preparation and the spectroscopic setup are provided in the Experimental Section). The signal corresponding to the 1s exciton (indicated by an arrow in \textbf{Figure}\,\ref{fig:fig1_abs_R_structure}b) is followed by a series of additional absorption peaks on the higher energy side, attributed to polaronic effects\cite{Dyksik2024polaron, thouin2019phonon, Neutzner2018exciton, urban2018non}. As a result, the optical transition associated with the 1s exciton is often obscured by a broad high-energy response\cite{Posmyk2022quantification}, which is particularly noticeable in reflectivity measurements where overlapping resonances are present (\textbf{Figure}\,\ref{fig:fig1_abs_R_structure}b). In general, the strongest signal observed in reflectivity is not always due to the 1s exciton, making a comparison with linear absorption crucial for accurately determining $E_{1s}$\cite{posmyk2023fine}.

Another important aspect when determining $E_b$ in 2D layered perovskites is the band gap energy $E_g$ ($E_b=E_g-E_{1s}$). In the 2D limit, one would expect a step-like feature associated with the band-to-band absorption\cite{Chernikov2014exciton, hansen2024measuring}. As presented in \textbf{Figure}\,\ref{fig:fig1_abs_R_structure}b the signal at $\sim 2.6$\,eV, related to the band-to-band absorption is significantly broadened, deviating from an expected step-like shape. The increased broadening is due to the higher-order excitonic states merging together with the quasi-particle band gap\cite{ziegler2022excitons, blancon2018scaling, Feldstein2020microscopic, movilla2023excitons}. Thus, applying Elliot's formula to determine the band gap energy $E_g$ in this context is challenging\cite{Neutzner2018exciton}. Specifically, the higher-order excitonic transitions are often disregarded when determining the band gap energy $E_g$ using the Elliot formalism, even though they may contribute to the step-like feature (\textbf{Figure}\,\ref{fig:fig1_abs_R_structure}b). Therefore, a robust and parameter-free approach for determining the band gap energy and thus the exciton binding energy $E_b$ in 2D layered perovskites is highly desirable.

Magneto-optical spectroscopy can be used to precisely determine the $E_g$ value using Landau level spectroscopy. Such a technique has been very successful in determining $E_g$ in the case of 3D perovskites\cite{miyata2015direct, yang2017unraveling, baranowski2020excitons} and recent studies have shown that it can be successfully applied to the case of 2D perovskites\cite{dyksik2020broad, dyksik2021tuning}. However, this approach requires samples with a high structural quality and/or the use of extreme magnetic fields to facilitate the observation of the Landau levels as the cyclotron frequency has to be higher than the scattering rate\cite{miura2008physics, Dyksik2022using}.

\begin{table}[b]
\centering
\setlength{\tabcolsep}{5pt}
\renewcommand{\arraystretch}{1.1}
\begin{tabular}{l|ccc}
&$E_g$ (eV) & $E_b$ (meV) & $c_{0}$ (\unit{\micro\electronvolt\per\tesla\squared}) \\ 
\hline
(BA)$_2$SnI$_4$ & \num{2.355 +- 0.022} & 210 & \num{0.45 +- 0.03}  
\\
(BA)$_2$PbI$_4$ - LT & \num{2.830 +-0.017} & 293 & \num{0.13}\cite{baranowski2019phase} 
\\
(BA)$_2$PbI$_4$ - HT & \num{2.646 +- 0.064} & 246 & \num{0.32}\cite{baranowski2019phase} 
\\
(PEA)$_2$SnI$_4$ & 2.084\cite{dyksik2020broad} & 174\cite{dyksik2020broad}& \num{0.68}\cite{dyksik2020broad} 
\\
(PEA)$_2$PbI$_4$ & 2.608\cite{dyksik2020broad} & 260 \cite{dyksik2020broad} & \num{0.36}\cite{dyksik2020broad} 
\\
(PEA)$_2$PbBr$_4$ & 3.4 \cite{Takagi2013influence} & 356 & \num{0.13+-0,02}
\\
(HA)$_2$PbI$_4$ & 2.6\cite{Tanaka2005Image} & 250\cite{Tanaka2005Image} & \num{0.33}\cite{baranowski2019phase} 
\\
(BA)$_2$PbBr$_4$ & 3.42\cite{Takagi2013influence} & 393\cite{Takagi2013influence} & \num{0.12}\cite{tanaka2005electronic} 
\\
(DA)$_2$PbI$_4$ & - & 320\cite{Ishihara1990Optical} & \num{0.16}\cite{baranowski2019phase}

\end{tabular}
\caption{Summary of the parameters obtained from magneto-optical spectroscopic studies. From the left: band gap energy $E_g$ (the definition of uncertainty for ratio method is presented in \textbf{Figure}\,\ref{Sfig:SI_uncertainty}), exciton binding energy $E_b$, diamagnetic coefficient $c_0$. The $c_0$ values for (PEA)$_2$PbBr$_4$ and (BA)$_2$SnI$_4$ are presented in \textbf{Figure}\,\ref{Sfig:SI_peapbbr4_Bfield} and \textbf{Figure}\,\ref{Sfig:SI_basni4_diamagnetyk}, respectively.
}
\label{tab:tab2}
\end{table}

Crucially, even if a full Landau fan chart\cite{miura2008physics} cannot be obtained, the spectral feature corresponding to the band gap onset can be clearly observed in the ratio spectrum measured in the magnetic field \emph{i.e.} ratio of the transmission in the field and zero-field spectrum. A typical ratio spectrum for (PEA)$_2$PbI$_4$ is presented in Fig.\,\ref{fig:fig1_abs_R_structure}c (in \textbf{Figure}\,\ref{Sfig:SI_bandgap_megagauss} for high-temperature (HT) phase of (BA)$_2$PbI$_4$ and (PEA)$_2$PbI$_4$). At low energies, complex variations in intensity are visible due to the shifts of the excitonic transitions\cite{baranowski2019phase} in a magnetic field. Around $\simeq 2.6$\,eV we observe a strong signal in the energy range of $E_g$. This additional component can be used to approximate the band gap energy (see \textbf{Figure}\,\ref{Sfig:SI_uncertainty}). Based on such a technique we estimate $E_g$ for 2D perovskite structure studied herein, including (BA)$_2$SnI$_4$ (\textbf{Figure}\,\ref{Sfig:SI_bandgap_basni4}), HT (BA)$_2$PbI$_4$ (\textbf{Figure}\,\ref{Sfig:SI_bandgap_megagauss}), low-temperature (LT) phase of (BA)$_2$PbI$_4$ (\textbf{Figure}\,\ref{Sfig:SI_bapbi4_spectra}), (PEA)$_2$PbI$_4$ (\textbf{Figure}\,\ref{Sfig:SI_bandgap_megagauss}). The determined $E_g$ values are summarized in Table\,\ref{tab:tab2}.

\subsection{Exciton binding energy scaling laws}

Due to the challenges in determining both $E_g$ and $E_{1s}$ from linear absorption spectra, the exciton binding energy $E_b$ for 2D layered perovskites might be subject to significant uncertainty. Therefore, we systematically investigate the family of 2D perovskites using a magneto-optical spectroscopic approach, enabling us to propose a simple $E_b$($E_g$) scaling law.

\begin{figure}
    \centering
    \includegraphics[width=1\linewidth]{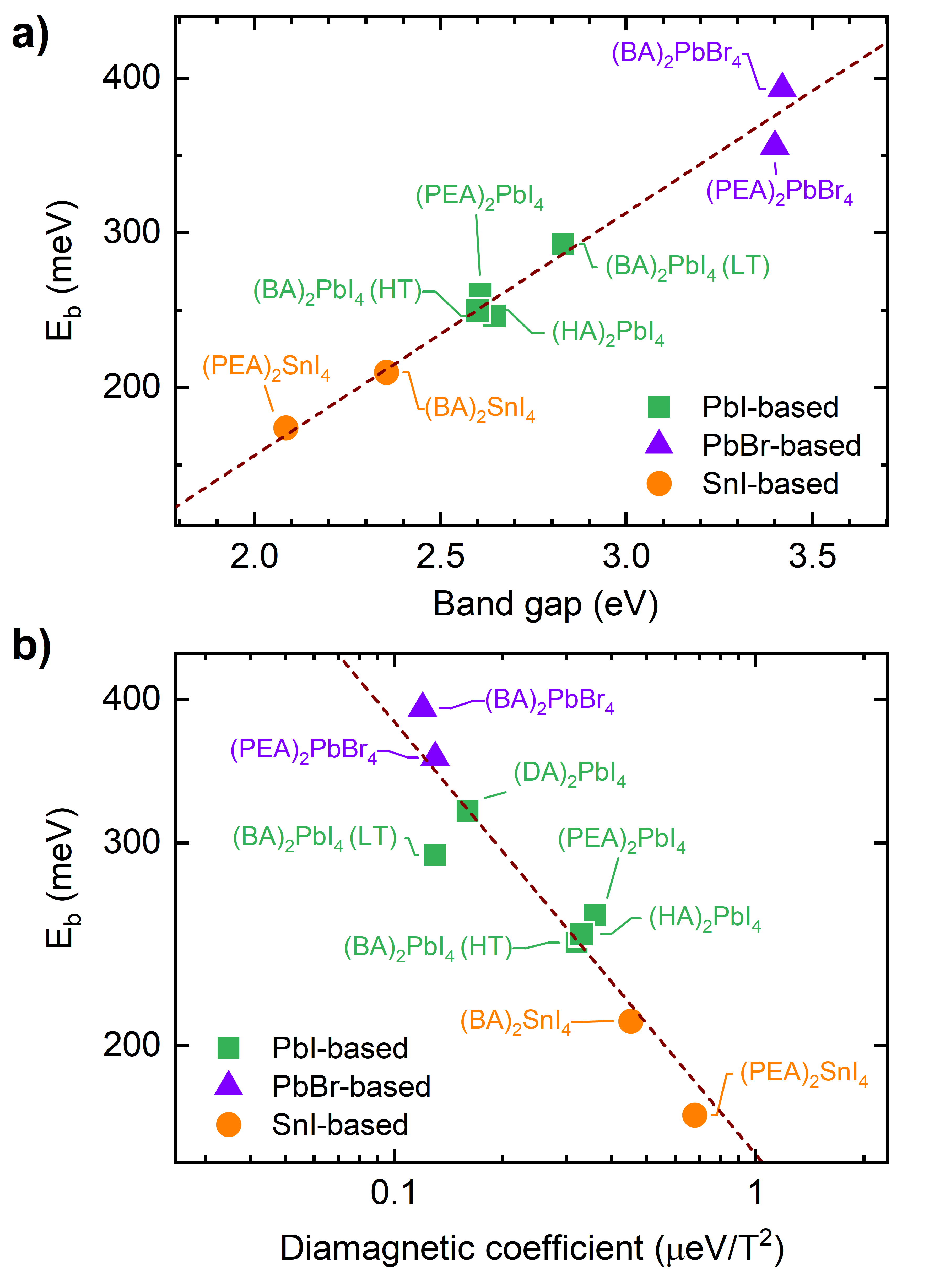}
    \caption{
    Exciton binding energy $E_b$ as a function of a) band gap $E_g$ and b) diamagnetic coefficient $c_0$. The dashed lines are fits. The source data is summarized in Table\,\ref{tab:tab2}.}
    \label{fig:fig2_parameters_vs_Eb}
\end{figure}

\textbf{Figure}\,\ref{fig:fig2_parameters_vs_Eb}a summarizes the band gap -- exciton binding energy dependence for different 2D layered perovskites. With the colour-coding, we differentiate between different metal-halide compositions. We notice that similar to fully inorganic ``classic'' semiconductors the exciton binding energy $E_b$ is a linear function of $E_g$\cite{Yu2010cardona,singh2007electronic,Baranowski2022exciton} (dashed line in \textbf{Figure}\,\ref{fig:fig2_parameters_vs_Eb}a). The fitting procedure yields a phenomenological law for the exciton binding energy $E_b=0.157 E_g-0.158$ (both $E_b$ and $E_g$ are in \unit{\electronvolt}.  It is interesting to note that the selection of organic spacer has only a marginal influence on the observed scaling between $E_b$ and $E_g$. The highly corrugated BA-based variants (in the low-temperature phase\cite{Takahashi2007tunable,Wong2017synthesis}) follow the same linear trend as structures with lower corrugation based on PEA (see Table\,\ref{tab:tab1}). As we will discuss later, this is not the case for the exchange interaction defining the exciton fine structure.  

The magneto-optical spectroscopy provides an alternative, indirect measure of the $E_b$ related to the so-called diamagnetic shift of the excitonic transition\cite{miura2008physics,baranowski2019phase, urban2020revealing, dyksik2021tuning, goryca2019revealing}. The diamagnetic coefficient of the 1s exciton $c_0$ is given by\cite{miura2008physics}: 
\begin{equation}
 c_0 = \frac{\mathrm{e}^2\langle r^2 \rangle}{8\mu}   
\label{eq:diamagnetyk}
\end{equation}
where $\mu$ is the reduced effective mass of the exciton and $\langle r^2 \rangle$ is the mean square expectation value of the wave function radius. Thus, $c_0$ contains information, about the exciton wave function extension, which is dependent on the exciton binding energy\cite{miura2008physics}. As we show in \textbf{Figure}\,\ref{fig:fig2_parameters_vs_Eb}b the $c_0$ is correlated with $E_b$ delivering a simple scaling law which effectively describes $E_b$ for a broad range of compositions. 
In the simple hydrogen model, the exciton binding energy $E_b$ scales like $\simeq c_0 ^{-\frac{1}{3}}$. By fitting the experimental data in \textbf{Figure}\,\ref{fig:fig2_parameters_vs_Eb}b (dashed line) we obtain $E_b=\num{161}\,c_0^{\num{-0.38}}$ 
where $E_b$ is in meV and $c_0$ in \unit{\micro\electronvolt\per\tesla\squared}. Interestingly, the obtained scaling law is relatively similar to the c$_0^{-\frac{1}{3}}$ dependency expected from the simple hydrogen model (3D or 2D).  

The determined scaling laws for $E_b$ have the following implications for the perovskite semiconductors; (i) we show that a very naive assumption of a linear scaling of $E_b$ and $E_g$ still holds for 2D layered perovskites regardless of steric effect. The obtained scaling law allows for a quick estimation of the excitonic binding energy $E_b$ if the band gap energy $E_g$ is known. The latter can be directly obtained from experimental techniques such as scanning tunnelling microscopy\cite{Zhang2022atomic}, time-resolved angle-resolved photoelectron spectroscopy\cite{Lee2021timeresolved}, or from band structure simulations. In general, the linear relationship between band gap and exciton binding energy is an empirical observation for fully inorganic semiconductors\cite{Baranowski2022exciton}, which also holds for transition metal dichalcogenides\cite{Jiang2017scaling} Moreover, our linear scaling law is consistent with the recent electro-absorption studies on 2D layered perovskites\cite{Hansen2023mechanistic}.  
(ii) We emphasize the obtained phenomenological $E_b(c_0)$ scaling law is rather surprisingly close to the dependency expected in a hydrogenic system although it is well-established that excitons in the 2D limit are subject to a complex dielectric screening \cite{Chernikov2014exciton,Feldstein2020microscopic, cudazzo2011dielectric} which is not taken into account in 3D or purely 2D hydrogen models. (iii) Most importantly, both scaling laws can serve as a benchmark for band structure theories for layered materials. The $E_g$ and $E_b$ values are typical outputs of such simulations, thus our simple scaling laws facilitate a quick approximation of excitonic effects. Furthermore, effective models, beyond DFT, are greatly anticipated, and phenomenological scaling laws derived herein can serve as a stimulus for future models. These will ultimately simplify the band structure description and facilitate further understanding of the photophysics of 2D layered perovskites. 

\section{Exciton fine structure in 2D layered perovskites}

\begin{figure}
    \centering
    \includegraphics[width=1\linewidth]{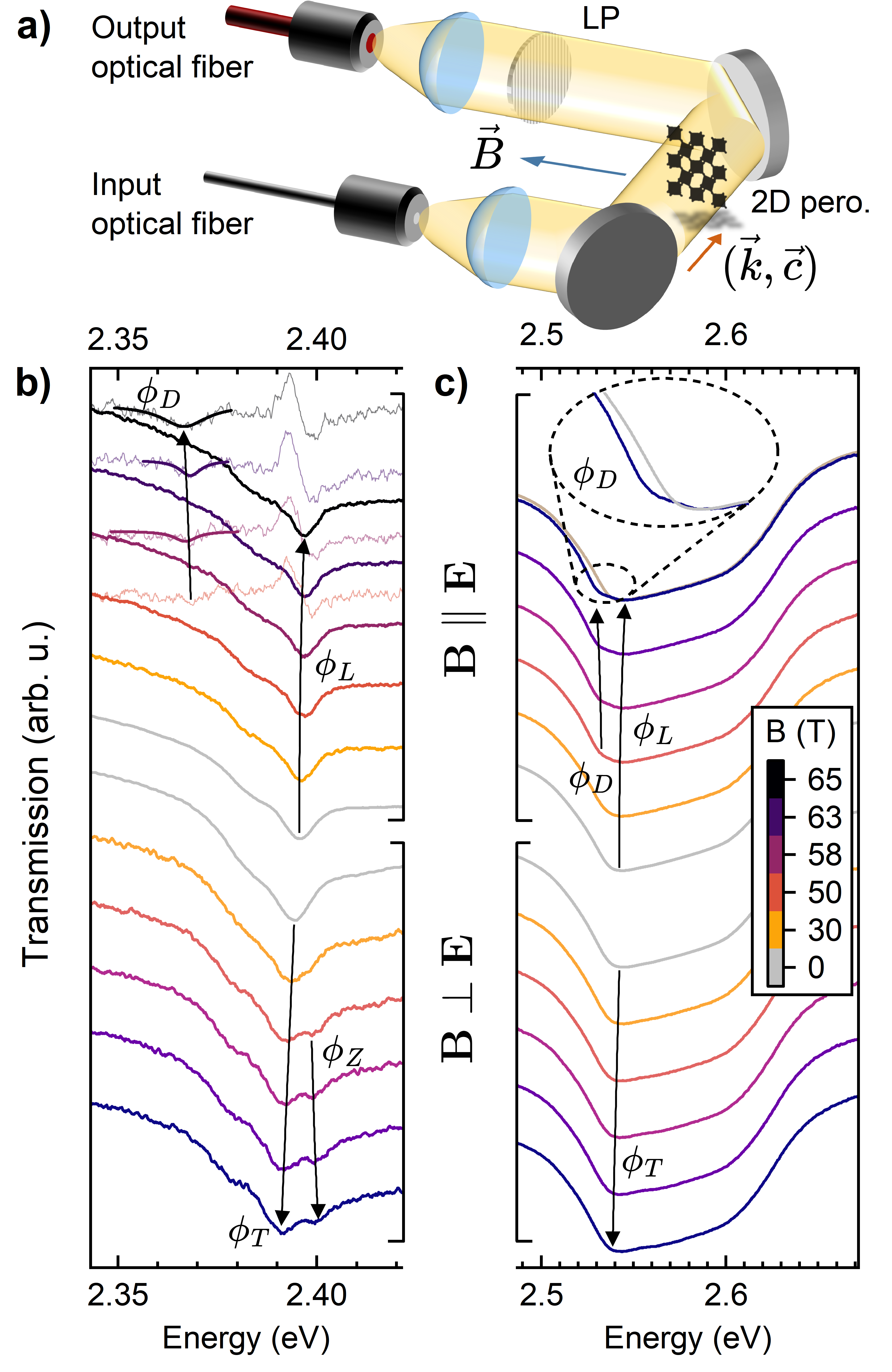}
    \caption{a) Scheme of the experiment in the Voigt geometry. The magnetic field vector \textbf{B} is normal to both the \textbf{k} vector of the probing light and \textbf{c} axis of the sample. LP stands for linear polarizer which selects either the $\textbf{B}\parallel \textbf{E}$ or $\textbf{B}\perp \textbf{E}$ configuration (see text). b) Transmission of HT (BA)$_2$PbI$_4$ measured in Voigt geometry for several magnetic field strengths. To better visualize the emergence of the brightened dark state at the low-energy range, we plot the ratio spectra (thin lines) for high-field data. c) Transmission of LT (BA)$_2$PbI$_4$ measured in Voigt geometry for several magnetic field strengths. The inset to panel c) shows an expanded view of the brightened dark state $phi_D$. In panels b) and c) labels $\phi_L$ and $\phi_T$ denote the longitudinal and transverse bright states, respectively, and $\phi_Z$ is the brightened \emph{gray} exciton. The arrows guide the eye showing the evolution of states with increasing magnetic field.}
    \label{fig:fig3_spectraBfield}
\end{figure}

Having established the scaling laws for the exciton binding energy $E_b$ in 2D layered perovskites, we turn our attention to the exciton fine structure -- a consequence of the different spin configurations of the electrons and holes within the excitonic quasiparticle. The resulting exciton fine structure splitting occurs on a much smaller energy scale compared to the Coulomb correlation. The exchange interaction that drives the splitting between excitonic states typically ranges from a few, to several tens of meV in 2D perovskites, while the $E_b$ is on the order of hundreds of meV\cite{blancon2018scaling,dyksik2021tuning}.  
Nevertheless, the excitonic fine structure has a significant impact on the optical response of a given material, and, most importantly, on their light emission efficiency\cite{canneson2017negatively,tamarat2019ground,xu2018long,Dyksik2021brightening,Thompson2024phononbottleneck}. It is important to highlight that the exchange interaction in 2D perovskites is several orders of magnitude larger than in ``classical'' III-V semiconductor nanostructures\cite{BenAich2020multiband}, leading to its crucial impact on their optical properties.

The band-edge excitons in 2D perovskites pair a hole from the top of the valence band (characterized with a total angular momentum j$^h=\frac{1}{2}$ and two possible projections on $z$-axis, j$_z^h=\pm\frac{1}{2}$)  with the low energy spin-orbit split-off conduction band with the same angular momentum (j$^e=\frac{1}{2}$, j$_z^e=\pm\frac{1}{2}$). Therefore, four band-edge exciton states are expected, three of them are optically active and one is dark ($\psi_D$)\cite{tanaka2005electronic,Quarti2024exciton}. The exchange interaction lifts the degeneracy between the bright and the dark states while the bright state degeneracy depends on the symmetry of the system. In the case of 2D perovskites, due to the low D$_{2h}$ symmetry, the degeneracy of bright excitonic states is lifted (\textbf{Figure}\,\ref{Sfig:SI_levels_inBfield_D2h})\cite{Posmyk2022quantification}. Thus, the optical response is driven by three bright excitonic states $\psi_X$, $\psi_Y$, and $\psi_Z$  which couple to the linearly polarized light along the crystallographic directions. The spacing, ordering, and degeneracy of the exciton manifold are scaled by the exciton binding energy and crystal lattice symmetry.\cite{BenAich2020multiband}. All these parameters can be controlled by selecting inorganic sublattice composition and/or organic spacers. Importantly, current reports indicate that in 2D perovskites the spacing within the fine structure is roughly 10 times larger\cite{Posmyk2022quantification,Dyksik2021brightening,Posmyk2024exciton, posmyk2024bright, zhang2023dark} than in 3D perovskites\cite{baranowski2019giant} or nanocrystals\cite{nestoklon2018optical,Yin2017excitonic,tamarat2019ground}. However, despite this fact a systematic investigation for a broad range of 2D perovskites is missing. Effectively a comprehensive description of the exciton fine structure is available only for (PEA)$_2$PbI$_4$\cite{Dyksik2021brightening,Posmyk2022quantification,Posmyk2024exciton}. The precise knowledge of energy splitting between the excitonic states and their ordering is crucial for operational devices, in principle, it is imperative to know the energetic position of both dark state $\psi_D$ and bright out-of-plane state $\psi_Z$, as these two do not efficiently couple with an in-plane electric field (\textbf{k}-vector normal to the 2D perovskite plane). 

\begin{figure*}
    \centering
    \includegraphics[width=1\linewidth]{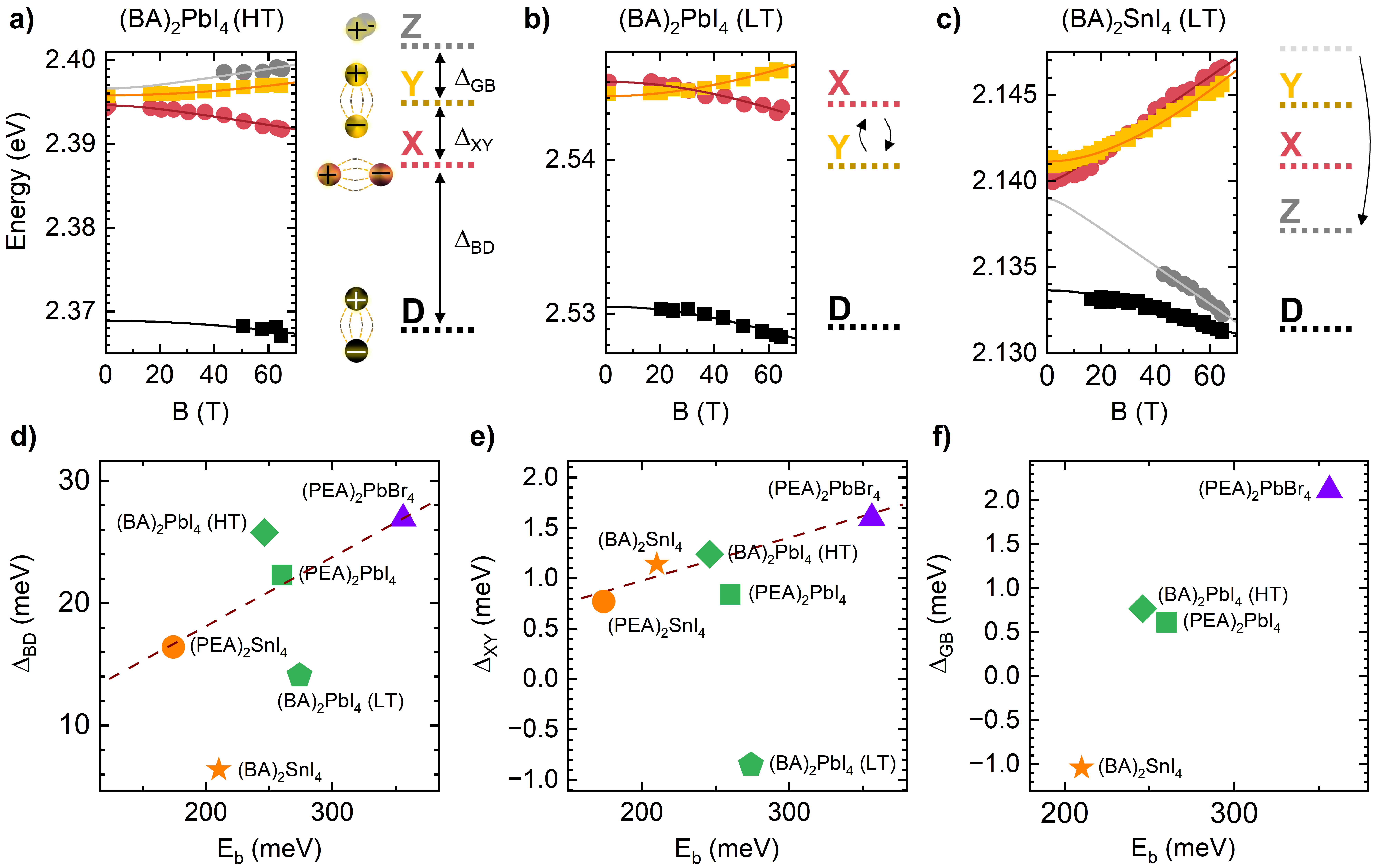}
    \caption{Evolution of optical transition energy for dark state and three bright states for a) HT (BA)$_2$PbI$_4$ b) LT (BA)$_2$PbI$_4$ and c) (BA)$_2$SnI$_4$. The inset to panels (a-c) depicts the energy structure of the exciton manifold for a given sample. d) The bright -- dark energy splitting ($\Delta_{BD}$) at $B=0$\,T in function of exciton binding energy $E_b$. e) The splitting between the in-plane excitonic states ($\Delta_{XY}$) in the function of $E_b$. f) The splitting between the out-of-plane excitonic state and the closest (in energy) bright in-plane state ($\Delta_{GB}$) in the function of $E_b$. The dashed line in panels d) and e) is a guide to the eye.}
    \label{fig:fig4_splittings}
\end{figure*}

In order to access all excitonic states, especially $\psi_Z$ and $\psi_D$ states, we perform magneto-optical spectroscopy in the Voigt geometry. As schematically presented in \textbf{Figure}\,\ref{fig:fig3_spectraBfield}a in this configuration we have $\mathbf{c} \parallel \mathbf{k} \perp \mathbf{B}$. When $B \neq 0$ the zero-field states $\psi_i$ ($i$=D,X,Y,Z) are no longer the eigen-states of the system. The bright in-plane states start to mix with $\psi_Z$ and $\psi_D$ making them optically active (\textbf{Figure}\,\ref{Sfig:SI_levels_inBfield_D2h}). Tracking the evolution of four states as a function of the magnetic field allows us to determine the energy of $\psi_Z$ and $\psi_D$ states, even though they are optically inactive at zero magnetic field\cite{yu2016effective}.

When $B>0$ the magnetic-field-induced $\phi$ states are considered as a solution to a simple eigen-problem $H\psi=\epsilon\phi$, where the Hamiltonian $H$ describes the interaction with the magnetic field in Voigt geometry (see Section\,\ref{Ssec:model} of Supplementary Information). In particular, two coupled pairs of states are formed; the longitudinal states ($\phi_L$ and $\phi_D$) which interact with light polarized along the magnetic field vector \textbf{B} ($\mathbf{E} \parallel \mathbf{B}$), and the transverse states ($\phi_T$ and $\phi_Z$) coupling to the light polarized perpendicular to the magnetic field vector ($\mathbf{E} \perp \mathbf{B}$). In the magnetic field, the energy $\epsilon$ of both pairs evolves as (Section\,\ref{Ssec:model} of Supplementary Information):

\begin{equation}    
\label{eq:shift1}
    \epsilon_{D,L}^{\vec{B}\parallel \vec{E}} (B) =  \dfrac{1}{2}\left(E_\text{D} +E_\text{L} \pm \sqrt{(E_\text{D} -E_\text{L})^2 + (g_L \mu_B B)^2} \right) 
\end{equation}  

\begin{equation}
 \label{eq:shift2}
     \epsilon_{Z,T}^{\vec{B}\perp \vec{E}} (B) =  \dfrac{1}{2}\left(E_\text{T} + E_\text{Z} \pm \sqrt{(E_\text{T} - E_\text{Z})^2 + (g_T \mu_B B)^2} \right) 
\end{equation}

We emphasize, that when $B \rightarrow 0$ the magnetic-field-induced $\phi_L$ and $\phi_T$ states transform into zero-field bright in-plane ones ($\psi_X$, $\psi_Y$). Furthermore, since the bright in-plane states are non-degenerate at 0\,T we note, in the above equations, the $E_L$ state represents either $E_X$ or $E_Y$, depending on the in-plane orientation of these bright states with respect to the magnetic field vector \textbf{B}. The same applies to the $E_T$ state.

In \textbf{Figure}\,\ref{fig:fig3_spectraBfield}b and c we show the evolution of the transmission spectra in the magnetic field for (BA)$_2$PbI$_4$ measured at \SI{2.2}{K}. We spectrally resolve two regions of interest, an optical transition at $\sim \SI{2.4}{\electronvolt}$ (\textbf{Figure}\,\ref{fig:fig3_spectraBfield}b) and $\sim 2.55$\,eV (\textbf{Figure}\,\ref{fig:fig3_spectraBfield}c), which are ascribed to excitonic transitions characteristic for the high-temperature phase (HT) and low-temperature phase (LT) of (BA)$_2$PbI$_4$, respectively. The ``freeze out'' of HT phase domains upon cooling has been previously observed and enables the study of both phases simultaneously (\textbf{Figure}\,\ref{Sfig:SI_bapbi4_spectra})\cite{baranowski2019phase,Yaffe2015excitons}. In particular, it allows us to observe the important impact of the phase transition on the exciton manifold as discussed below.       

In order to independently access the longitudinal and transverse states we perform transmission measurements in the $\mathbf{E} \parallel \mathbf{B}$ and $\mathbf{E} \perp \mathbf{B}$ linear polarizations, respectively, as schematically shown in \textbf{Figure}\,\ref{fig:fig3_spectraBfield}a. 
In the longitudinal configuration for both HT (\textbf{Figure}\,\ref{fig:fig3_spectraBfield}b) and LT (\textbf{Figure}\,\ref{fig:fig3_spectraBfield}c) phases we observe the longitudinal bright state $\phi_L$ blue shifting with increasing magnetic field strength. When the magnetic field is larger than $\approx 40$\,T, an additional optical transition labelled $\phi_D$ emerges at the low energy side of the $\phi_L$ bright exciton. This new optical transition gains in oscillator strength and red shifts with increasing magnetic field and is associated with a brightened dark excitonic state $\phi_D$. In the transverse configuration ($\mathbf{E} \perp \mathbf{B}$), we observe a transverse bright state $\phi_T$ red shifting with the increasing magnetic field $\mathbf{B}$ for both phases (\textbf{Figure}\,\ref{fig:fig3_spectraBfield}b and c). Moreover, in the HT phase at a high magnetic field, a new blue shifting state emerges on the high energy side of $\phi_T$ that is attributed to the brightened $\phi_Z$ state. Unfortunately, due to the large spectral broadening, this state is not resolved in the LT phase. We obtain quantitatively similar spectra for the LT phase of the tin-iodide variant \emph{i.e.} (BA)$_2$SnI$_4$. The spectra for both $\mathbf{E} \parallel \mathbf{B}$ and $\mathbf{E} \perp \mathbf{B}$ geometries are presented in \textbf{Figure}\,\ref{Sfig:SI_basni4_Bfieldspectra}. 

\begin{table*}
\centering
\setlength{\tabcolsep}{3pt}
\renewcommand{\arraystretch}{1.2}
\begin{tabular}{l|ccccccc}
& $\beta$ (\unit{\degree}) & $\delta$ (\unit{\degree}) & $\Delta_{BD}$ (meV)& $\Delta_{XY}$ (meV) & $\Delta_{GB}$ (meV) & $\sqrt{\langle r^2 \rangle}$ (nm) & $I$ (meV) \\ 
\hline
(BA)$_2$SnI$_4$ & \num{14.2 +- 0.1}\cite{Takahashi2007tunable,Wong2017synthesis} & \num{12.9 +- 0.1}\cite{Takahashi2007tunable,Wong2017synthesis} & \num{6.4 +- 1.1}  & \num{1.1 +- 1.1} & \num{-1.0 +- 1.2} & \num{1.31} & \num{9.7}      
\\
(BA)$_2$PbI$_4$ - LT & \num{15.6 +-0.3}\cite{Billing2007Synthesis,Menahem2021strongly} & \num{12.9 +- 0.1}\cite{Billing2007Synthesis,Menahem2021strongly} & \num{14.1 +- 1.1} & \num{-0.85 +- 0.12} & - & \num{0.86} & \num{21.1}$^{\alpha}$         
\\
(BA)$_2$PbI$_4$ - HT & \num{12.5 +- 0.1}\cite{Billing2007Synthesis,Mitzi1996Synthesis,Menahem2021strongly} & \num{5.7+-0.1}\cite{Billing2007Synthesis,Mitzi1996Synthesis,Menahem2021strongly} & \num{25.8 \pm 0.5} & \num{1.24 +- 0.78} & \num{0.77 +- 0.81} & \num{1.14} & \num{40.3}     
\\
(PEA)$_2$SnI$_4$ & \num{12.6+-0.9}\cite{Takahashi2007tunable,Gao2019Highly,Park2019Highly} & \num{1.4+-0.1}\cite{Takahashi2007tunable,Park2019Highly} & \num{17.9 +- 1.1}\cite{Dyksik2021brightening} & \num{0.77+-0.1} & - & \num{1.5} & \num{26.8}$^{\alpha}$                  
\\
(PEA)$_2$PbI$_4$ & \num{13.6 +- 0.5}\cite{Menahem2021strongly,du2017two} & \num{3.1+-1}\cite{Menahem2021strongly,du2017two} & \num{22.3 +- 3.3} & \num{0.8+-1.3} & \num{0.61 +- 1.32} & \num{1.2} & \num{34.6}                  
\\
(PEA)$_2$PbBr$_4$ & \num{14.6+-0.6}\cite{Shibuya2009,Ma2017single} & \num{3.9+-0.4}\cite{Shibuya2009,Ma2017single} & \num{27.0 +- 2.1} & \num{1.6 +- 0.2} & \num{2.1 +- 0.2} & \num{0.97} & \num{43.1}     
\end{tabular}
\caption{Summary of structural data and fine structure parameters. From the left: $\beta$ distortion angle (defined in \textbf{Figure}\,\ref{fig:fig5_theory}c), $\delta$ corrugation angle (defined in \textbf{Figure}\,\ref{fig:fig5_theory}d), bright--dark splittings ($\Delta_{BD}$), bright--bright splitting ($\Delta_{XY}$), bright--gray splitting ($\Delta_{GB}$), 1s exciton wavefunction extension $\sqrt{ \langle r^2 \rangle }$, exchange energy $I$, $^{\alpha}$estimated upper limit (see main text). $\beta$ and $\delta$ angles refer to the averaged values based on the referenced reports. $\Delta_{XY}$ value for (PEA)$_2$SnI$_4$ is determined in \textbf{Figure}\,\ref{Sfig:SI_peasni_polPL}.
}
\label{tab:tab1}
\end{table*}

The analysis of the magneto-optical response of (BA)$_2$PbI$_4$ (at both phases) and (BA)$_2$SnI$_4$ reveals a rich landscape of possible ordering of the excitonic states, which is presented in \textbf{Figure}\,\ref{fig:fig4_splittings}a-c. The solid lines in \textbf{Figure}\,\ref{fig:fig4_splittings}a-c are fits with eqs.\,\ref{eq:shift1}-\ref{eq:shift2}.  
For all investigated compounds, the lowest laying dark state is followed by the bright triplet. However, the ordering of the bright states evidently changes between different phases or for different metal cations. For HT (BA)$_2$PbI$_4$ the lowest-lying dark state is followed by two bright states with the in-plane dipole moments ($\psi_X$ and $\psi_Y$) whereas the excitonic state with the out-of-plane dipole moment ($\psi_Z$) is the highest-lying component. Such an order is also found in (PEA)$_2$PbI$_4$ and (PEA)$_2$PbBr$_4$ compounds\cite{Dyksik2021brightening, Posmyk2024exciton}. 
In addition, we observe that the two in-plane excitonic states reverse their order between the HT and LT phases of (BA)$_2$PbI$_4$ (cf. \textbf{Figure}\,\ref{fig:fig4_splittings}a and b). We emphasize that both phases are monitored under the same experimental conditions (see Fig\,\ref{Sfig:SI_bapbi4_spectra}). The reversed order of $\psi_X$ and $\psi_Y$ excitonic states reflects that the distortion of the metal-halide octahedra units (which differentiates the two phases\cite{Billing2007Synthesis}) has an important impact on the exciton fine structure. Although we were not able to observe the $\psi_Z$ state in the LT phase of (BA)$_2$PbI$_4$, the magnetic-field-coupled bright transverse $\phi_T$ state red shifts with increasing magnetic field. 
Therefore, the second state from the pair, $\phi_Z$, should be positioned higher, making it the highest energy state in the exciton manifold of the LT (BA)$_2$PbI$_4$. Finally, in the case of (BA)$_2$SnI$_4$ (\textbf{Figure}\,\ref{fig:fig4_splittings}c) we find that the $\phi_Z$ state is pushed below the bright in-plane states, following the theoretical predictions for 3D perovskites\cite{yu2016effective}.

\begin{figure*}
    \centering
    \includegraphics[width=1\linewidth]{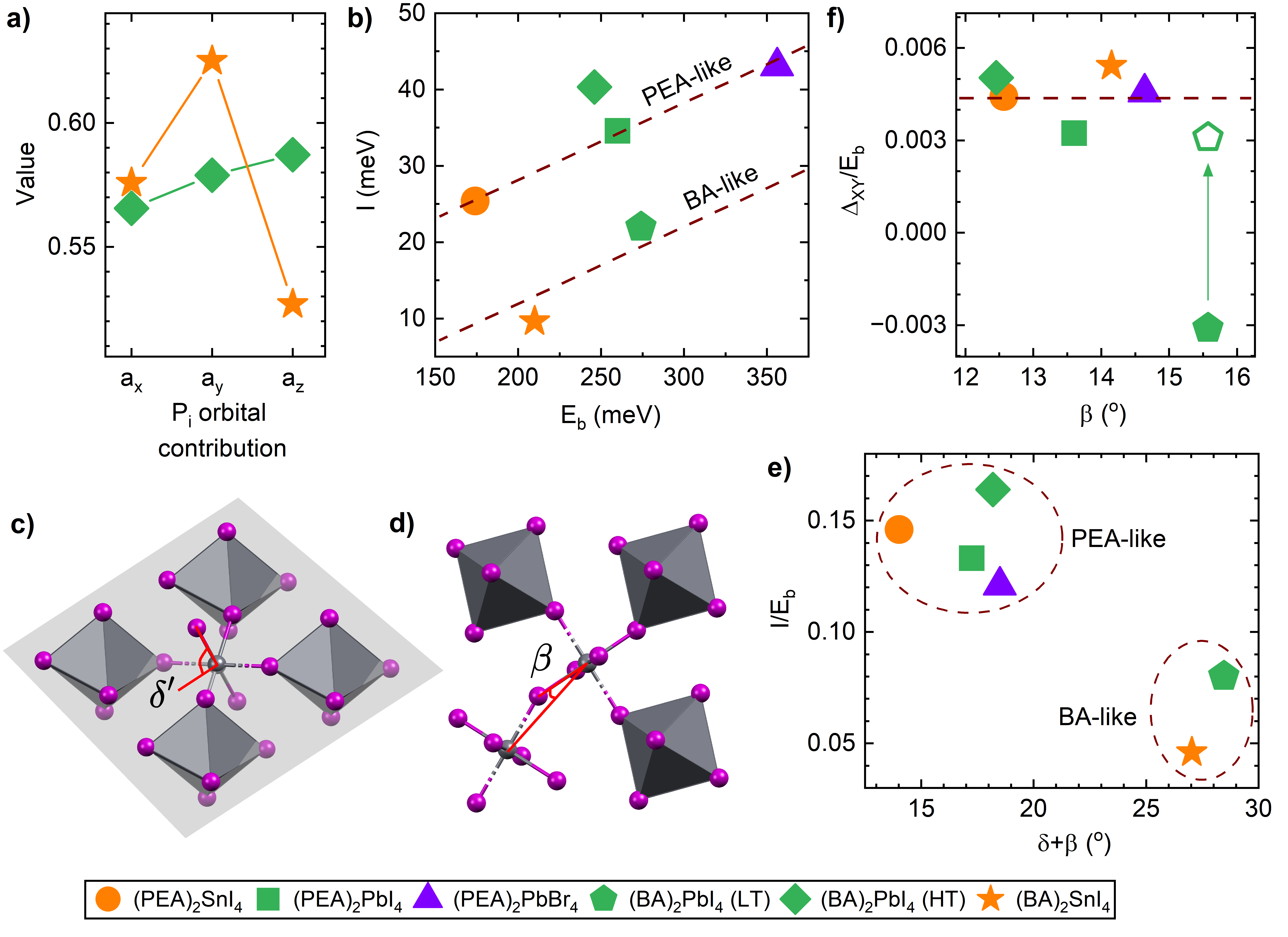}
    \caption{
    a) The calculated contribution $a_i$ of the $P_i$ ($i=x,y,z$) orbital to the conduction band. b) The exchange energy $I$ in the function of exciton binding energy $E_b$. Dashed lines are guides to the eye. c) The out-of-plane corrugation angle of the inorganic framework $\delta$ is defined as $\delta = \SI{90}{\degree} - \delta^{'}$, where $\delta^{'}$ angle is measured between the inorganic plane (gray color) and vector connecting the Pb atom and the apical I.
    d) The in-plane distortion $\beta$ angle here defined as the deviation of the bridging angle Pb-I-Pb from \SI{180}{\degree}.  
    e) The ratio of exchange energy $I$ and $E_b$ in the function of sum of distortion $\beta$ and corrugation $\delta$ angles. 
    f) The ratio of the in-plane states splitting and $E_b$ in the function of $\beta$ angle.
    }
    \label{fig:fig5_theory}
\end{figure*}

In \textbf{Figure}\,\ref{fig:fig4_splittings}d-f we summarize the exciton manifold parameters $\Delta_{BD}$ (the energy difference between the lowest-lying in-plane bright state and the dark exciton), $\Delta_{XY}$ (the energy splitting between the bright in-plane states) and $\Delta_{GB}$ (the energy splitting between the $\psi_Z$ state - \emph{gray} exciton - and the in-plane bright state) as a function of the exciton binding energy (for reference splittings are defined in \textbf{Figure}\,\ref{fig:fig4_splittings}a).  In general, we observe a similar trend for the different splittings in the exciton manifold -- the $\Delta_{BD}$, $\Delta_{XY}$ $\Delta_{GB}$ increase with the exciton binding energy $E_b$. Such behaviour is directly related to the reduction of the exciton Bohr radius (with increased $E_b$) and the related enhancement in the electron-hole exchange interaction. Although the dashed line in \textbf{Figure}\,\ref{fig:fig4_splittings}d,e is drawn simply as a guide to the eye, most of the data points follow a linear trend. The absolute value of $\Delta_{XY}$ for the LT (BA)$_2$PbI$_4$ equals to $\sim 1$\,meV and also follows the expected trend. The only exceptions are observed for $\Delta_{BD}$ (\textbf{Figure}\,\ref{fig:fig4_splittings}d) in the low-temperature motif of BA-based compounds, (BA)$_2$SnI$_4$ and (BA)$_2$PbI$_4$. We postulate the reduced $\Delta_{BD}$ for these compounds is a hallmark of the importance of lattice corrugation on the exciton fine structure. Indeed, one can notice that all PEA-based compounds and HT (BA)$_2$PbI$_4$ share similar crystal structures \emph{i.e.} similar distortion and corrugation of the inorganic sublattice\cite{Billing2007Synthesis,Menahem2021strongly,du2017two} as summarized in Table\,\ref{tab:tab1}. Simultaneously, the LT phases of BA-based compounds strongly contrast with the remaining compounds\cite{Billing2007Synthesis,Takahashi2007tunable,Wong2017synthesis}. In this context, it is worth noting that the $\Delta_{GB}$ is similar for (PEA)$_2$PbI$_4$ and HT (BA)$_2$PbI$_4$ (cf. \textbf{Figure}\,\ref{fig:fig4_splittings}f and \textbf{Figure}\,\ref{Sfig:SI_pea_bright_shift}) and the order of the excitonic states ($\psi_Z$ above the $\psi_X$ and $\psi_Y$) is preserved in (PEA)$_2$PbBr$_4$. Only for (BA)$_2$SnI$_4$ we observe the reordering of $\psi_Z$ and the bright in-plane states. In Table\,\ref{tab:tab1} we summarize all the exciton fine structure parameters. The $g$-factors obtained from the analysis of the magneto-optical spectroscopic data are presented in Table\,\ref{Stab:SI_gfactor}.

\section{Discussion}

In order to deliver deeper insights into the observed ladder of excitonic states we use an effective mass model\cite{Yu2018Oscillatory,Thompson2024phononbottleneck,fu2017neutral,yu2016effective} to describe the exciton fine structure. Diagonalization of the exchange Hamiltonian leads to the following parameterization of the exciton energy landscape (Section\,\ref{Ssec:model} of Supplementary Information):
\begin{equation}
\begin{aligned}
        E_D &= 0 \quad(\psi_D) \\
    E_X&=  2 a_x^2 I  \quad(\psi_X) \\
    E_Y&=  2 a_y^2 I  \quad(\psi_Y)\\
    E_Z&=  2 a_z^2 I  \quad(\psi_Z)
    \label{eq:manifold}
\end{aligned}
\end{equation}
where $I$ is the exchange integral (hereafter referred to as the exchange energy) determining the energy scale of the exciton fine structure. The $a_x$, $a_y$, $a_z$ parameters indicate the contribution of the $P_x$, $P_y$, $P_z$ orbitals to the conduction band edge states and obey the following normalization:
\begin{equation}
   a_x^2+a_y^2+a_z^2=1
   \label{eq:norm}
\end{equation}
The $a_i$ coefficients depend on the symmetry of the crystal, determining the ordering of the bright states and splitting. For instance, for cubic symmetry all $a_i$ coefficients are equal $\frac{1}{\sqrt{3}}$\cite{fu2017neutral, ramade2018fine} and bright states are degenerate. In such a case the $\Delta_{BD}$ is a $\frac{2}{3}$ of the exchange energy $I$. In the case of reduced symmetry of the lattice, all coefficients are different and the degeneracy of the bright excitonic states is lifted. In general, the $a_i$ coefficients are complex functions of spin-orbit coupling and crystal field\cite{yu2016effective,Yu2018Oscillatory, ramade2018fine}, however, the detailed analysis of them in this context is beyond the scope of this work. Knowing the energy of all four excitonic states it is possible to extract the exchange energy $I$ and $a_i$ coefficients based on the above equations. 
In \textbf{Figure}\,\ref{fig:fig5_theory}a we present the
$P_i$ orbital contribution to the conduction band-edge state extracted from the exciton manifold of HT (BA)$_2$PbI$_4$ and (BA)$_2$SnI$_4$ (according to eq.\,\ref{eq:manifold}). 
One can notice that the more corrugated structure of (BA)$_2$SnI$_4$\cite{Takahashi2007tunable,Wong2017synthesis} is characterized by a larger variation of $a_i$ coefficients. Notably, $a_z$ has the smallest value for (BA)$_2$SnI$_4$ pushing the $\psi_Z$ state below the in-plane bright states (\textbf{Figure}\,\ref{fig:fig4_splittings}c). 

In principle, the exchange energy $I$ scales with the exciton binding energy $E_b$ or the exciton Bohr radius\cite{fu1999excitonic,BenAich2020multiband}. To demonstrate the scaling for 2D layered perovskites in \textbf{Figure}\,\ref{fig:fig5_theory}b we plot the exchange energy $I$ as a function of $E_b$. For the two samples, LT (BA)$_2$PbI$_4$ and (PEA)$_2$SnI$_4$, for which we are missing information about $\psi_Z$ state, we estimate the upper limit of $I$ as a $3/2\Delta_{BD}$ (eq.\,\ref{eq:manifold}). 

We observe that $I$ values are grouped into two sets (\textbf{Figure}\,\ref{fig:fig5_theory}b) following either the \emph{PEA-like} or \emph{BA-like} trend.
The former set is characterized by relatively low distortion $\beta$ and corrugation $\delta$ angles of the inorganic framework, defined in \textbf{Figure}\,\ref{fig:fig5_theory}c and d, respectively ($\beta$ and $\delta$ angles are summarized in Table\,\ref{tab:tab1}). The highly distorted compounds in the \emph{BA-like} set (LT (BA)$_2$PbI$_4$ and (BA)$_2$SnI$_4$) show reduced $I$ values. In both sets the exchange energy $I$ can be approximated by a similar linear trend of $E_b$, while the dependency for highly-distorted compounds is down-shifted.

Based on \textbf{Figure}\,\ref{fig:fig5_theory}b we note the distortion/corrugation of the inorganic framework plays a crucial role in determining the exchange energy $I$. To better visualize this effect in \textbf{Figure}\,\ref{fig:fig5_theory}e we show the $I/E_b$ ratio as a function of the sum of corrugation and distortion angles ($\beta + \delta$). We observe that the increasing distortion/corrugation of the crystal lattice reduces the relative value of exchange energy $I$ to the exciton binding energy $E_b$ (\textbf{Figure}\,\ref{Sfig:SI_individual_exchange} shows the I/$E_b$ plot vs. individual distortion angles). Therefore we reveal that the steric effect has a significant impact on the exchange energy $I$. Importantly, the expectation that $\Delta_{BD}$ scales with $E_b$ is only valid for compounds of a similar crystal structure. Our experimental findings are consistent with recent work by Quarti et al. on ab initio modeling of the exciton fine structure in 2D perovskites\cite{Quarti2024exciton}. Their detailed theoretical analysis highlights the influence of lattice distortions on the exciton fine structure. Notably, they conclude that significant lattice distortions can dramatically alter the fine structure, potentially making the bright exciton the lowest-energy excitation.

At this point, we comment on the $\Delta_{XY}$ splitting and its dependence on distortion angles. In \textbf{Figure}\,\ref{fig:fig5_theory}f we show the $\Delta_{XY}/E_b$ ratio plotted in a function of $\beta$ angle. All studied samples exhibit a similar $\Delta_{XY}/E_b$ ratio (including the absolute value of LT-(BA)$_2$PbI$_4$, indicated by an arrow). A constant $\Delta_{XY}/E_b$ value versus the in-plane distortion angle $\beta$ suggests that the linear $\Delta_{XY}$ dependence on the exciton binding energy $E_b$ (Fig.\,\ref{fig:fig4_splittings}e) is driven solely by the increasing $E_b$. Analogous trends in $\Delta_{XY}/E_b$ ratio is observed as a function of the $\delta$ corrugation angle (\textbf{Figure}\,\ref{Sfig:SI_Dxy_delta}).

A similar conclusion on the importance of the steric effect on $\Delta_{BD}$ can be drawn by investigating the exchange energy $I$ dependence on the exciton Bohr radius $a_B$. In fully-inorganic ``traditional'' semiconductors such as GaAs, CdTe, the exchange energy $I$ scales as $\simeq \frac{1}{a_B^3}$\cite{fu1999excitonic}. Traditionally, the $a_B$ concerns material systems where the hydrogenic model applies and is defined as the maximum in the radial probability density. With magneto-optical spectroscopy, we can probe the root mean square (rms) of the 1s exciton wave-function extension $\sqrt{\langle r^2 \rangle}$ which is proportional to $a_B$. For instance, in the 2D hydrogen model, the $\sqrt{\langle r^2 \rangle}$ equals to $\sqrt{6}a_B$\cite{Zipfel2018spatial}. The $\sqrt{\langle r^2 \rangle}$ determines the diamagnetic shift of the excitonic transition \emph{i.e.} the quadratic part of the exciton energy shift in the magnetic field $\Delta E=c_0B^2$\cite{miura2008physics}, where $c_0$ is defined by eq.\,\ref{eq:diamagnetyk}. 

\begin{figure}
    \centering
    \includegraphics[width=1\linewidth]{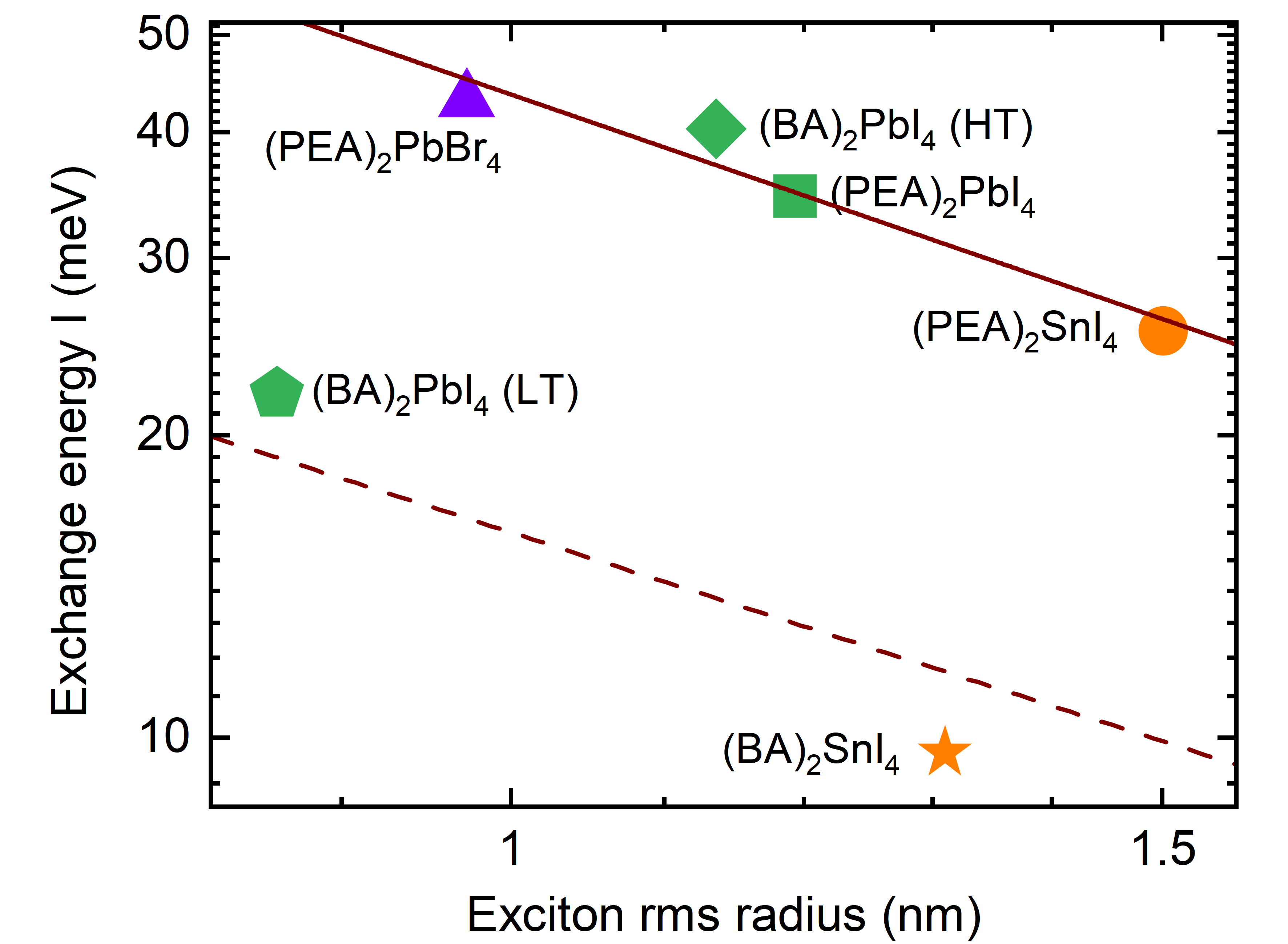}
    \caption{The exchange energy $I$ in the function of 1s exciton rms radius. The dashed line is the least-square-fit to the data points.}
    \label{fig:fig6_bohr_radius}
\end{figure}

Based on the diamagnetic coefficient $c_0$ of the 1s exciton (summarized in Table\,\ref{tab:tab2}) and the reduced effective mass (Table\,\ref{Stab:SI_masses}) we calculate the 1s exciton rms radius. In \textbf{Figure}\,\ref{fig:fig6_bohr_radius} we plot the exchange energy $I$ as a function of the 1s exciton rms radius. The dashed line shows the $I\sim\frac{1}{r^{-b}}$ dependency with the determined scaling factor $b=\num{1.18 +- 0.27}$. This simple formula effectively describes the exchange integral $I$ behaviour in \emph{PEA-like} compounds. Simultaneously, we see that the same renormalized scaling law describes well the two representatives of strongly corrugated compounds. Thus the generalized prediction of the exchange energy evolution is valid only for the families of 2D perovskites with similar lattice corrugation and distortion, underscoring the importance of the steric effect. We emphasize the determined scaling law is rather closer to $\frac{1}{a_B}$ than $\frac{1}{a_B^3}$ expected for 3D semiconductors\cite{fu1999excitonic} which can be attributed to the different dimensionality of layered perovskite system. 

In conclusion, magneto-optical spectroscopy has allowed us to probe essential electronic parameters of 2D layered perovskites, such as band gap energy, exciton binding energy, and the exciton fine structure. This work represents the first comprehensive study across a broad range of 2D perovskites with different organic sublattices, metal, and halide compositions. We determine a linear scaling relation between the exciton binding energy and the band gap, indicating that 2D perovskites share similarities in this aspect with many other semiconducting systems. Importantly, the scaling between exciton binding energy and the bandgap holds regardless of the choice of the organic spacer, which controls the distortions of the metal-halide framework. Therefore, this relationship should also be applicable for estimating exciton binding energy in structures subjected to external pressure or strain, even though these stimuli affect lattice distortions\cite{tu2019probing, jiao2021strain, liu2018isothermal, liu2019manipulating, zhang2019pressure}.
Furthermore, we quantified how the exciton fine structure evolves as a function of exciton binding energy, revealing that templating with organic spacer provides an effective control over exchange interactions and the resulting splitting between bright and dark excitonic states. Such a unique feature of 2D perovskites provides a potential strategy for optimizing emission properties of 2D perovskites by engineering the bright--dark exciton splitting through steric effects, opening new avenues for tailoring their optoelectronic performance. 

\section{Experimental Section}
\subsection{Sample synthesis details}

All chemicals were used without further purification. Solutions and thin films were prepared inside a nitrogen-filled glove box. Quartz substrates were sonically cleaned with a soapy solution, deionized water, acetone, and isopropanol. The surface was subsequently treated through UV–ozone for 20 min and thin films were cast a maximum of 15 min after the final treatment step.

A 2:1 molar ratio of BAI, PEAI, PEABr (($>$98\% - Sigma-Aldrich) to SnI$_2$ PbI$_2$ or PbBr$_2$ salts were dissolved in DMSO ($>$99.8\% Alfa Aesar) to prepare 0.5 M precursor solutions. The solutions were ready after 3 h of constant stirring at room temperature. The PEA(BA)$_2$Pb(Sn)I(Br)$_4$ thin films were spin-coated on top of 1 cm$^2$ quartz substrates in a single-step process at 5000 rpm for 60s. Toluene or chlorobenzene was used as an antisolvent and dripped during the spin-coating process. The films were annealed at 80-100 $^\circ$C for 10 min on a hotplate.

\subsection{Optical spectroscopy in high magnetic field}

\emph{Non-destructive pulsed magnets: }optical spectroscopy measurements in the high magnetic field were performed using a nitrogen-cooled pulsed magnet capable of generating a maximum magnetic field of $B \simeq 65$\,T, with pulse duration of approximately $\sim 200$\,ms. The sample was placed in a helium cryostat at the center of the magnetic field. Unless otherwise noted, all data was collected with the sample in pumped liquid helium at $T=2.2$\,K. A tungsten halogen lamp served as the broad-band white light source. White light was delivered to the sample via an optical fiber, and the transmitted light was collected using a lens coupled to a second fiber. The transmitted light was then dispersed and detected through a monochromator with a diffraction grating and a nitrogen-cooled CCD camera. The linear polarization was resolved \emph{in situ} using a broadband polarizer. 
The measurements were conducted either in the Voigt configuration, where the sample’s $\mathbf{c}$-axis was perpendicular to the magnetic field vector \textbf{B} and parallel to the light’s $\mathbf{k}$ vector (\textbf{Figure}\,\ref{fig:fig3_spectraBfield}a), or in Faraday geometry ($\mathbf{c} \parallel \mathbf{k} \parallel \mathbf{B}$). 

\emph{Megagauss magnetic fields:} A single-turn coil technique was employed to generate pulsed destructive magnetic fields of up to $B=140$\,T with a typical pulse duration of \SI{8}{\micro s}. During the pulsed magnetic field generation process, the strong electromagnetic force destroys the single-turn coil. The samples were placed in a specially designed liquid-He flow-type cryostat, where the temperature can be controlled from 5\,K to room temperature\cite{Yang2021heavy}. The temperature of the sample was monitored by a type-E thermocouple (nickel-chromium/constantan) and the value of the magnetic field was measured by a calibrated pick-up coil wound close to the sample. A xenon arc-flash lamp was used as the light source. The light was guided and collected by 800-\unit{\micro m}-diameter optical fibers. 

\subsection{Model of the exciton fine structure}
We employ a microscopic many-particle theory \cite{kira2006many, Feldstein2020microscopic, Thompson2024phononbottleneck} to derive the excitonic energy landscape and optical selection rules \cite{Thompson2024phononbottleneck}].  Following previous work \cite{Thompson2024phononbottleneck}, we derive the excitonic Hamiltonian in the exchange basis giving access to the energy states and optical dipole given the orbital character of the electron and hole states.  Full parametrisation of each material structure is beyond the scope of this work, particularly the exchange constant and orbital composition, which we extract from our theoretically derived equations for the energy levels. More details and the magnetic field dependence can be found in the Supplementary Information (Section\,\ref{Ssec:model}). \newline

\textbf{Acknowledgements:}

M.D. acknowledges support from the National Science Centre Poland within the SONATA grant (2021/43/D/ST3/01444). P.P. acknowledges the National Science Centre, Poland grant no.\ 2020/38/A/ST3/00214. M.B. acknowledges support from the National Science Centre Poland within the OPUS
LAP grant (2021/43/I/ST3/01357). This study has been partially supported through the EUR grant NanoX no. ANR-17-EURE-0009 in the framework of the ``Programme des Investissements d’Avenir''.
E.M. acknowledges funding from the Deutsche Forschungsgemeinschaft (DFG) via SFB 1083 and the regular project 504846924.
J.J.P.T.  acknowledges support from a EPSRC Programme Grant [EP/W017091/1].

\bibliography{Biblio}

\end{document}


\title{Supplementary Information\\ Steric Engineering of Exciton Fine Structure in 2D Perovskites}

\author{Mateusz Dyksik}
\email[]{mateusz.dyksik@pwr.edu.pl}
\affiliation{Department of Experimental Physics, Faculty of Fundamental Problems of Technology, Wroclaw University of Science and Technology, Wroclaw, Poland}

\author{Michal Baranowski}
\affiliation{Department of Experimental Physics, Faculty of Fundamental Problems of Technology, Wroclaw University of Science and Technology, Wroclaw, Poland}

\author{Joshua J. P. Thompson}
\affiliation{Department of Materials Science and Metallurgy, University of Cambridge, Cambridge CB3 0FS, United Kingdom}

\author{Zhuo Yang}
\affiliation{The Institute for Solid State Physics, The University of Tokyo, Kashiwanoha 5-1-5, Kashiwa, Chiba, Japan}

\author{Martha Rivera Medina}
\affiliation{Zernike Institute for Advanced Materials, University of Groningen, Nijenborgh 4, 9747 AG Groningen, The Netherlands}


\author{Maria Antonietta Loi}
\affiliation{Zernike Institute for Advanced Materials, University of Groningen, Nijenborgh 4, 9747 AG Groningen, The Netherlands}

\author{Ermin Malic}
\affiliation{Department of Physics, Philipps-Universit\"{a}t Marburg, Renthof 7, 35032 Marburg}

\author{Paulina Plochocka}
\email{paulina.plochocka@lncmi.cnrs.fr}
\affiliation{Department of Experimental Physics, Faculty of Fundamental Problems of Technology, Wroclaw University of Science and Technology, Wroclaw, Poland}
\affiliation{Laboratoire National des Champs Magn\'etiques Intenses, EMFL, CNRS UPR 3228, University Grenoble Alpes, University Toulouse, University Toulouse 3, INSA-T, Grenoble and Toulouse, France}

\keywords{}
                             
\maketitle
\clearpage
\section{SUPPLEMENTARY FIGURES}

\begin{figure*}[h!]
    \centering
    \caption{\textbf{Band gap energy of HT (BA)$_2$PbI$_4$ determined from spectroscopy at MegaGauss magnetic fields.}\newline
    a) Transmission ratio spectra of HT (BA)$_2$PbI$_4$ for selected magnetic field strengths. The signal due to band gap E$_g$ and 1s exciton is indicated by arrows. b) For reference, the same experiment is performed on an analogous sample (PEA)$_2$PbI$_4$, for which both energies are known\cite{dyksik2020broad,dyksik2021tuning}}
    \includegraphics[width=0.7\linewidth]{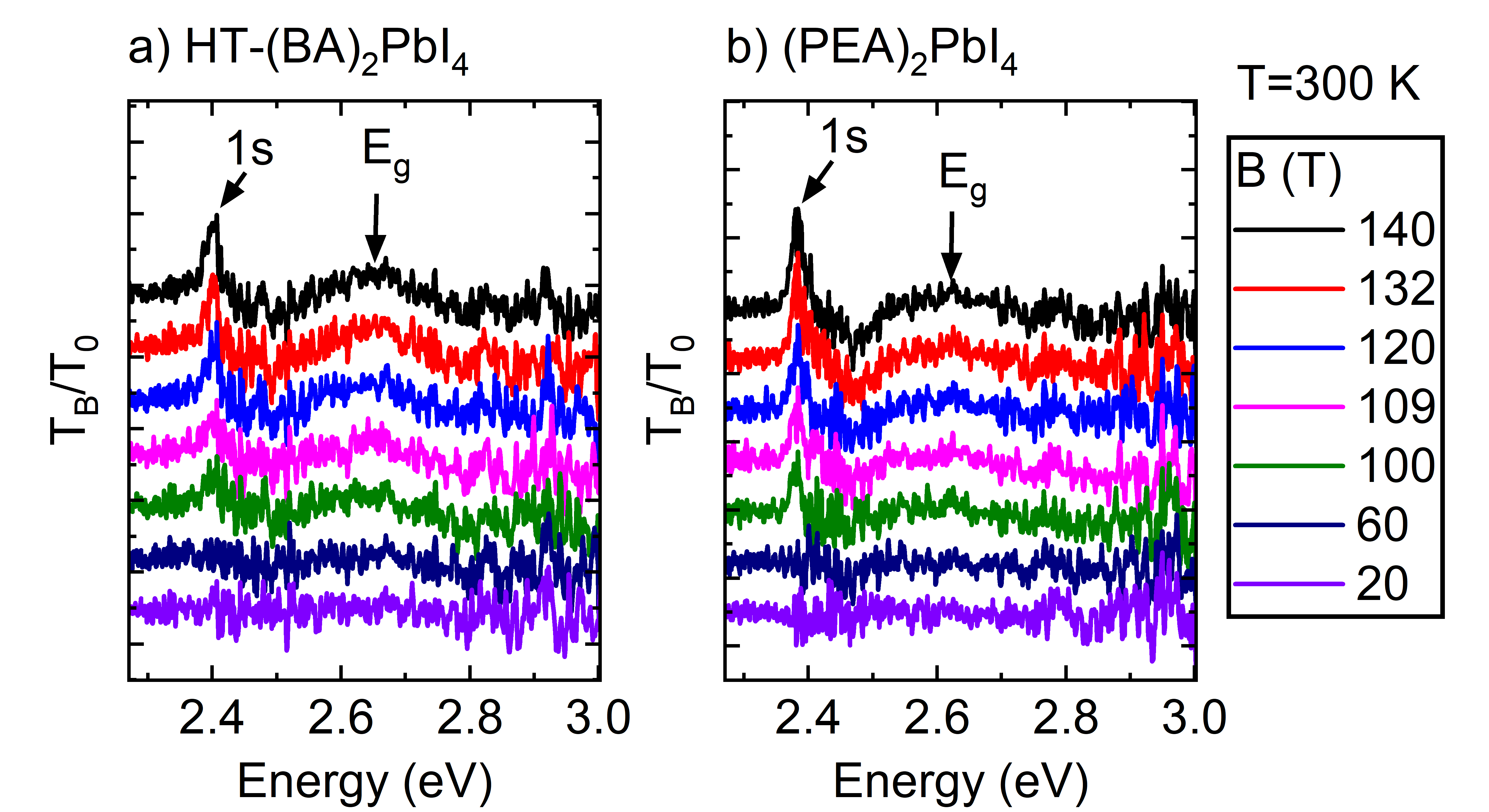}
\label{fig:SI_bandgap_megagauss}
\end{figure*}

\begin{figure*}[h!]
    \centering
        \caption{\textbf{Band gap of (BA)$_2$SnI$_4$.}\newline
        Transmission ratio spectra of (BA)$_2$SnI$_4$ for selected magnetic field strengths. The band gap energy E$_g$ and 1s exciton energy are indicated by arrows.
        }
        \includegraphics[width=0.6\linewidth]{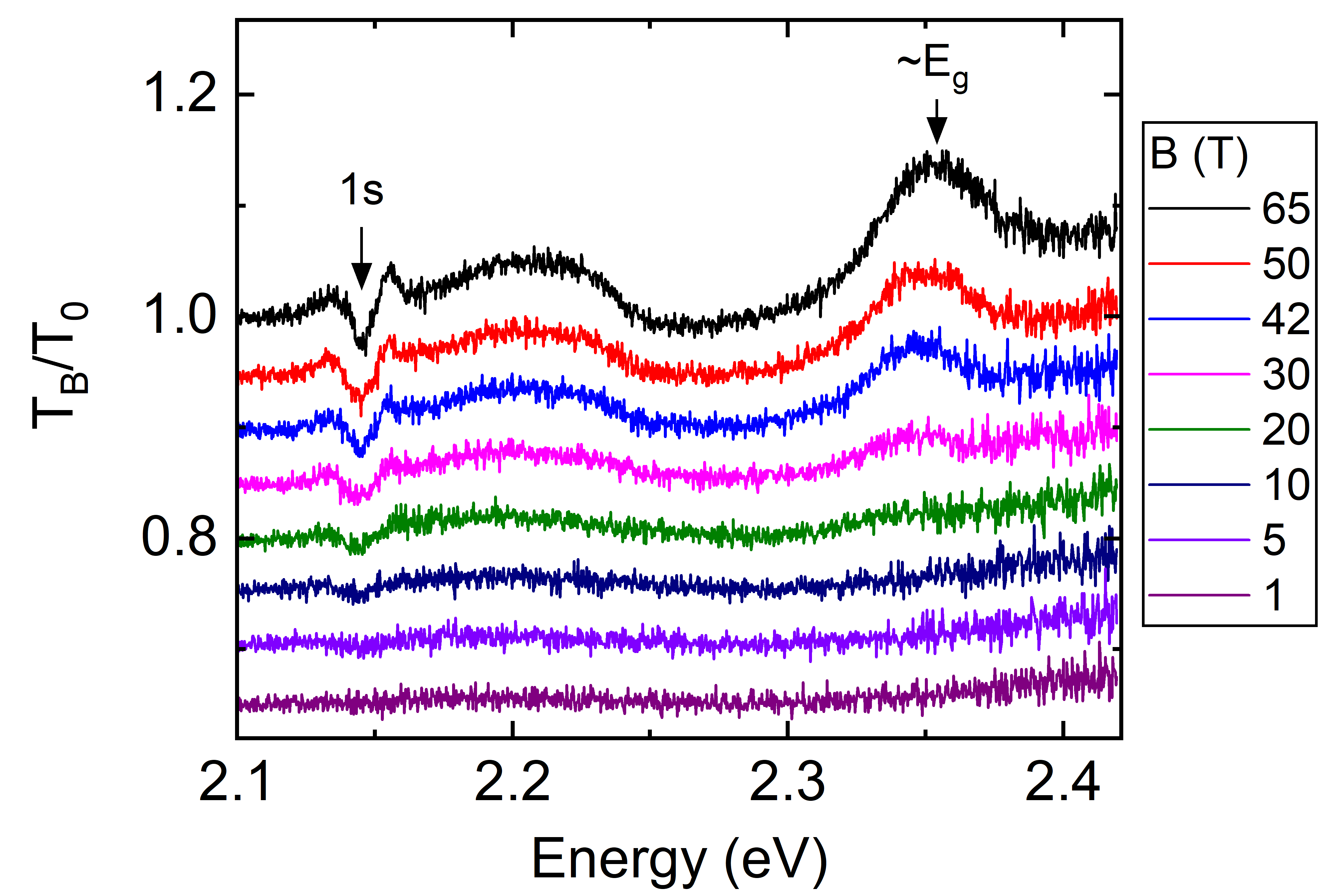}
\label{fig:SI_bandgap_basni4}
\end{figure*}

\begin{figure}[h!]
    \centering
        \caption{\textbf{Low temperature spectra of (BA)$_2$PbI$_4$.}\newline
        Low-temperature absorbance spectra of (BA)$_2$PbI$_4$ showing the signal due to both HT and LT phases. In red the transmission ratio spectrum is shown (transmission measured at B=65\,T divided by zero-field spectrum) in the spectral range of E$_g$. A clear resonance is observed, allowing for the quasi-particle band gap determination.  
        }
    \vspace{0.2cm}
    \includegraphics[width=0.6\linewidth]{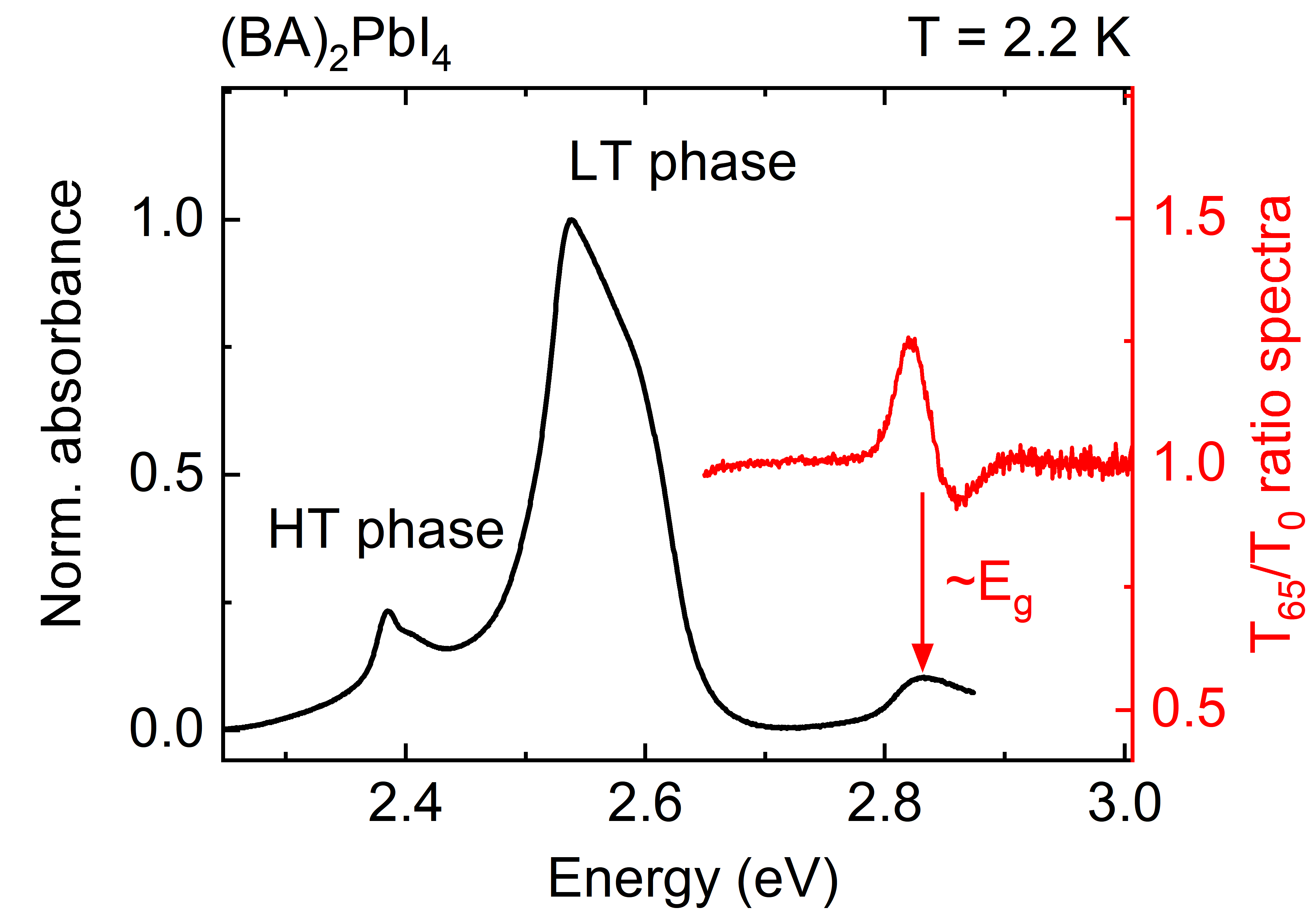}
    \label{fig:SI_bapbi4_spectra}
\end{figure}

\begin{figure*}[h!]
    \centering
        \caption{\textbf{Determination of band gap energy and its uncertainty from the ratio spectrum.}\newline
        a) Transmission ratio spectrum of LT (BA)$_2$PbI$_4$ (transmission measured in the magnetic field divided by zero-field spectrum) for selected field strengths, in the spectral range of the band gap energy E$_g$. The inflection point, indicated by dashed line, approximates the band gap energy.
        b) The uncertainty u(E$_g$) is determined as in the Figure. The same approach is used for other samples studied in the current work, summarized in Table\,\ref{Otab:tab2} of the main text.}
        \vspace{0.2 cm}
        \includegraphics[width=0.9\linewidth]{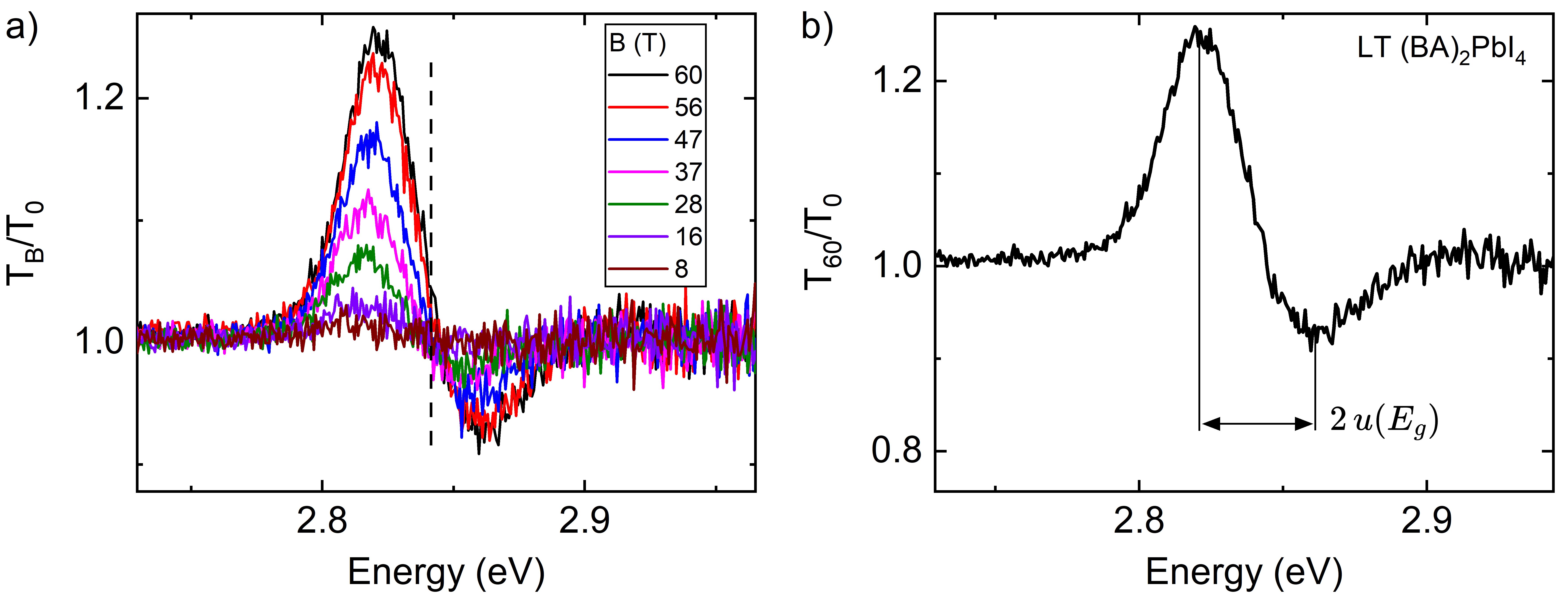}
    \label{fig:SI_uncertainty}
\end{figure*}

\begin{figure}[h!]
    \centering
    \caption{\textbf{Diamagnetic shift in (PEA)$_2$PbBr$_4$.}\newline 
    a) Transmission spectra of (PEA)$_2$PbBr$_4$ measured for two circular polarizations $\sigma +$ and $\sigma -$ for selected strengths of magnetic field. 
    b) The energy shift $\Delta E_{1s}=\frac{E_{\sigma +}+E_{\sigma -}}{2}$ of the 1s exciton ($\simeq 3.05$\,eV) in the function of magnetic field. Solid line stand for $\Delta E_{1s}=c_0 B^2$. The fitting yields $c_0 = \SI{0.13 +- 0.02}{\micro \eV \per\tesla\squared}$. c) Zeeman splitting with a $g$-factor of \num{1.3 +- 0.1}.    
    }
\vspace{0.2cm}
\includegraphics[width=1\linewidth]{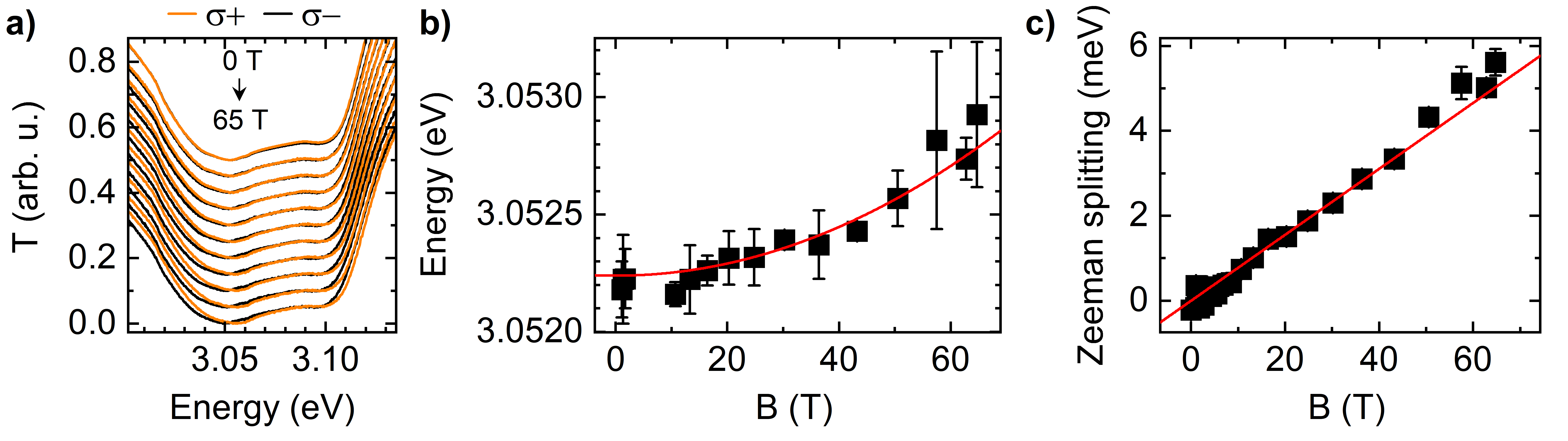}
\label{fig:SI_peapbbr4_Bfield}
\end{figure}

\begin{figure}[h!]
    \centering
    \caption{\textbf{Diamagnetic shift in (BA)$_2$SnI$_4$.}\newline 
    a) Transmission spectra of (BA)$_2$SnI$_4$ measured at \num{0} and \SI{65}{\tesla}.
    b) The energy shift $\Delta E_{1s}$ of 1s exciton in the magnetic field. Solid line stand for $\Delta E_{1s}=c_0 B^2$ dependence. The determined c$_0$ equals to \SI{0.45 +- 0.03}{\micro \eV \per\tesla\squared}
    }
    \vspace{0.2 cm}
    \includegraphics[width=0.9\linewidth]{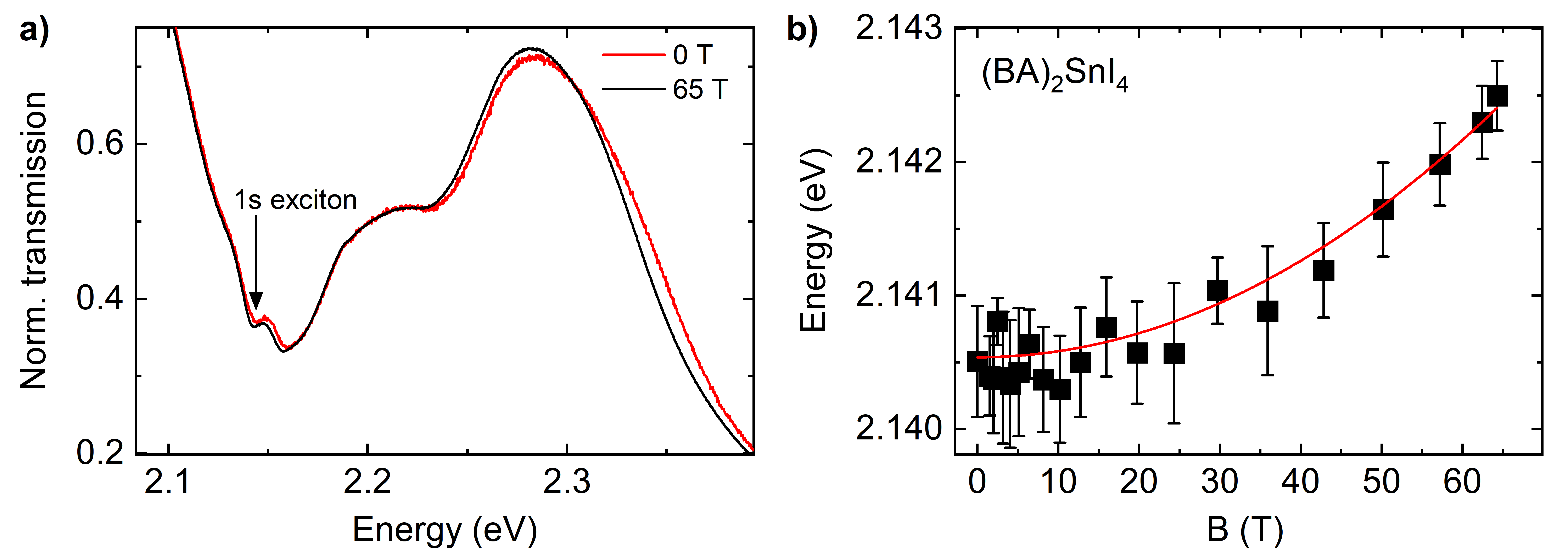}
    \label{fig:SI_basni4_diamagnetyk}
\end{figure}

\begin{figure}[h!]
    \centering
    \caption{\textbf{Fine structure splitting of the band-edge excitons for 2D (D$_{2\textrm{h}}$) symmetry.}\newline
    The optical selection rules, allowing to access the respective states, are indicated (\textbf{E} is light electric field vector and \textbf{c} is the crystallographic axis perpendicular to quantum well slab). At $B=0$\,T (left panel), $|g\rangle$ - the ground state (no exciton); $|\psi_D\rangle$ - dark state. The $|\psi^X\rangle$ and $|\psi^Y\rangle$ are bright states with in-plane dipole moment and $|\psi_Z\rangle$ state is a bright state with out-of-plane dipole moment. At $B>0$ and $\mathbf{B}\perp\mathbf{k}\parallel\mathbf{c}$ (right panel) all four states ($|\phi_D\rangle$, $|\phi_Z\rangle$, $|\phi_{L}\rangle$, $|\phi_{T}\rangle$) have nonzero dipole moment in the plane of 2D perovskite.
    }
    \includegraphics[width=0.85\linewidth]{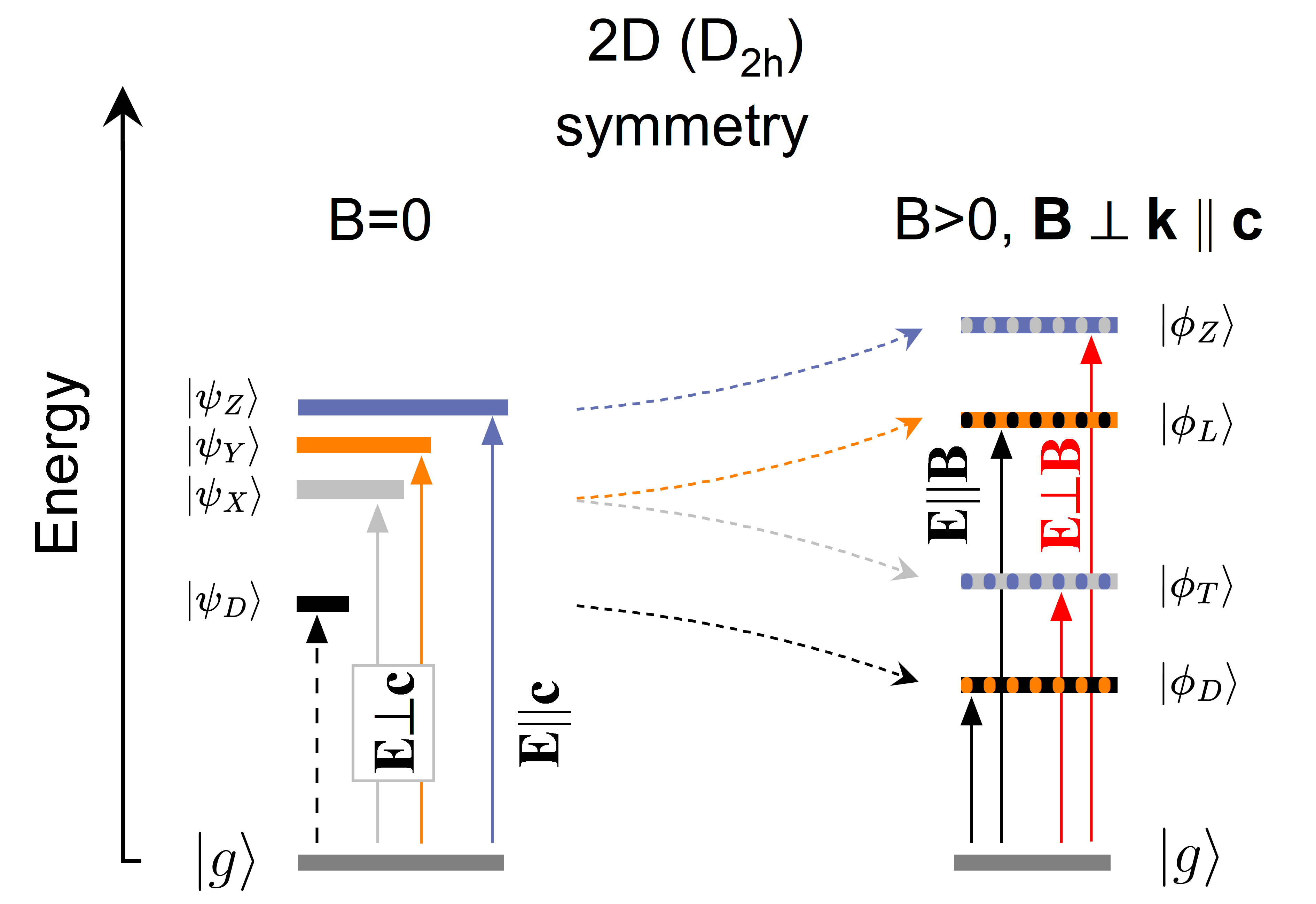}
    \label{fig:SI_levels_inBfield_D2h}
\end{figure}

\begin{figure}[h!]
    \centering
    \caption{\textbf{Transmission of (BA)$_2$SnI$_4$ in Voigt geometry.}\newline
    2nd derivative of transmission of (BA)$_2$SnI$_4$ measured in Voigt geometry for several magnetic field strengths for a) $\textbf{B}\perp \textbf{E}$ and b) $\textbf{B}\parallel \textbf{E}$ configurations. In the high magnetic field (bottom curves) all four excitonic states are observed. 
    }
    \includegraphics[width=0.7\linewidth]{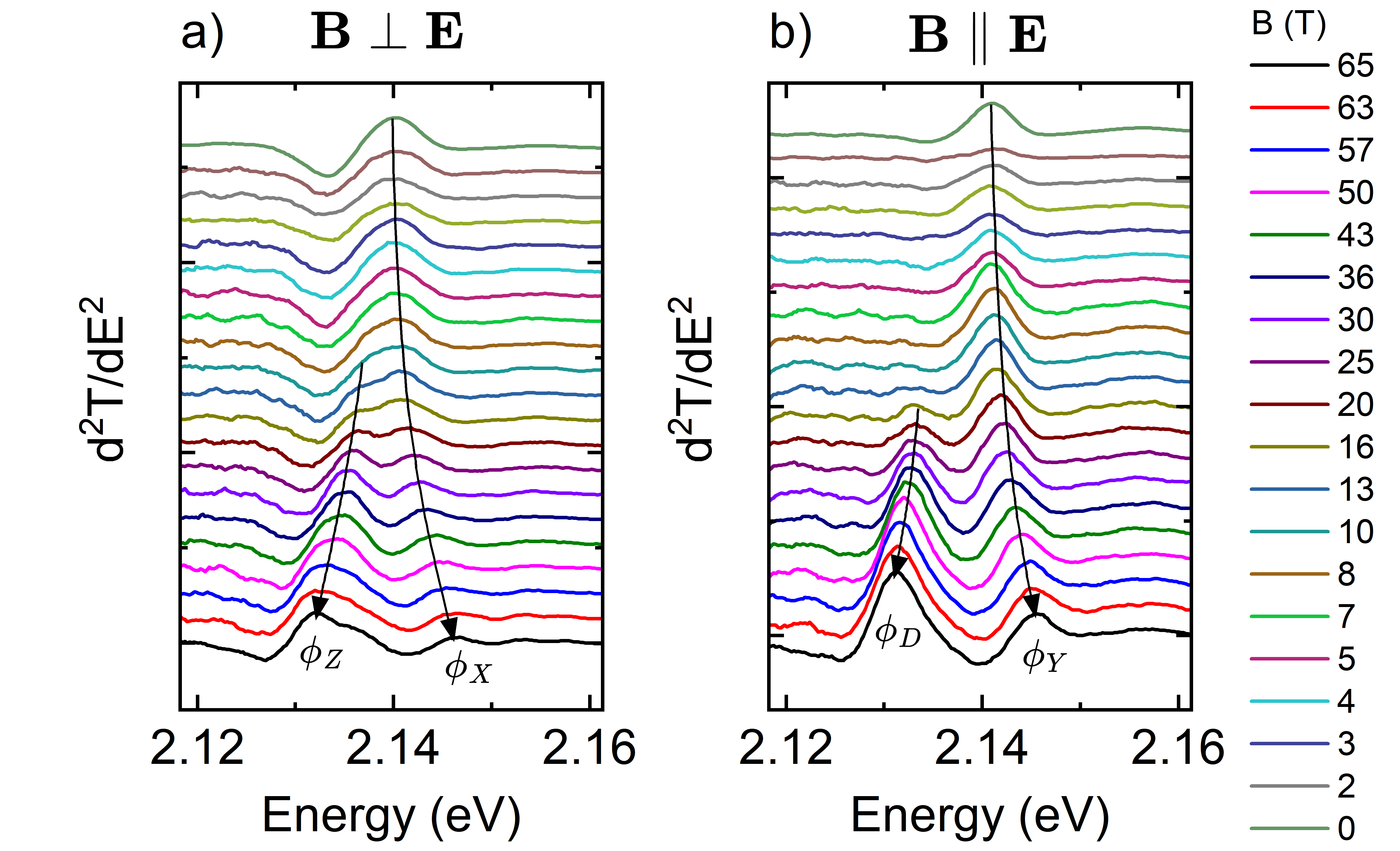}
    \label{fig:SI_basni4_Bfieldspectra}
\end{figure}

\begin{figure*}[h!]
    \centering
    \caption{\textbf{The energy shift of bright excitons in (PEA)$_2$PbI$_4$ and (PEA)$_2$PbBr$_4$}\newline
    The evolution of optical transition energy for three bright states of (PEA)$_2$PbI$_4$ and (PEA)$_2$PbBr$_4$. Solid lines stand for fits with eq.\,\ref{Oeq:shift1}-\ref{Oeq:shift2} of the main text. 
    }
    \vspace{0.2 cm}
    \includegraphics[width=1\linewidth]{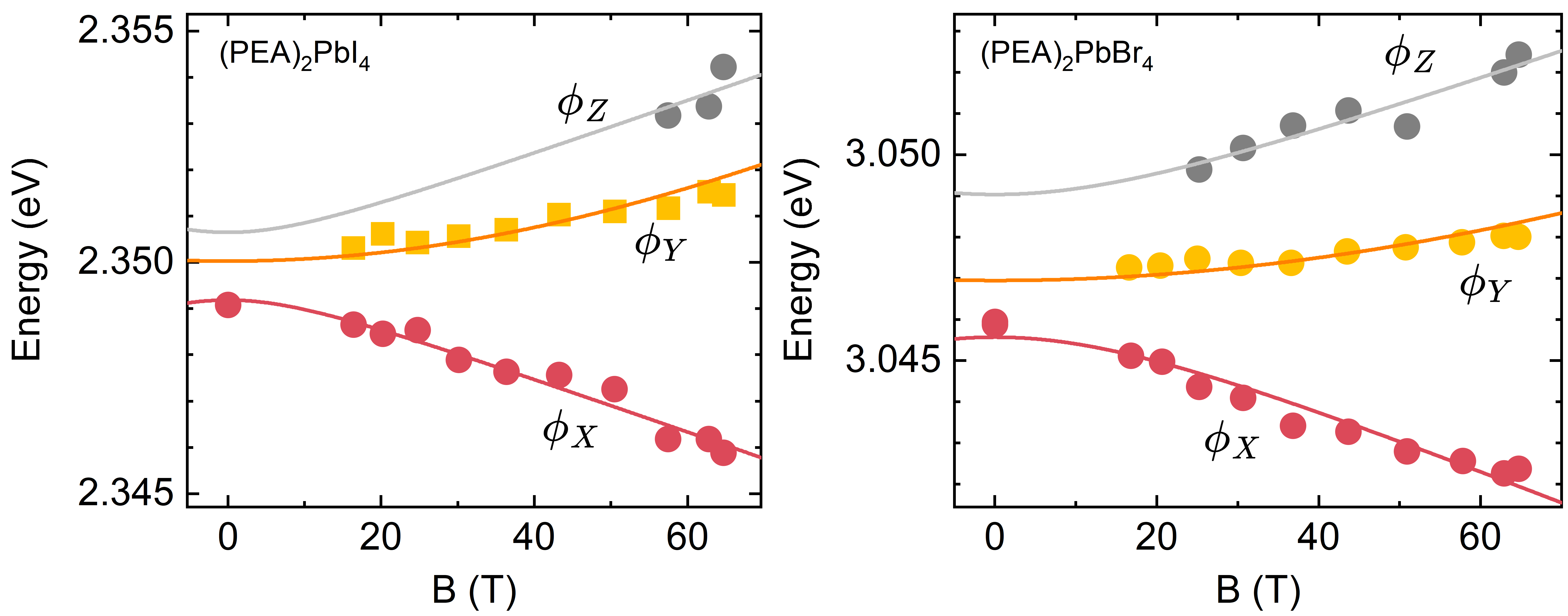}
    \label{fig:SI_pea_bright_shift}
\end{figure*}

\begin{figure*}[h!]
    \centering
    \caption{\textbf{$\Delta_{XY}$ in (PEA)$_2$SnI$_4$.}\newline
    In order to approximate the energy splitting between the bright in-plane states of (PEA)$_2$SnI$_4$ we performed photoluminescence (PL) measurements in the linear basis at T=\SI{4.2}{K}. a) The PL spectra for two orthogonal polarizations. A clear difference between the \SI{0}{\degree} and \SI{90}{\degree} curves is visible. b) The PL peak energy in the function of polarization angle. The solid line is a cosine fit, yielding the $\Delta_{XY}$=\SI{0.77 +- 0.1}{m\electronvolt}. 
    }
    \includegraphics[width=1\linewidth]{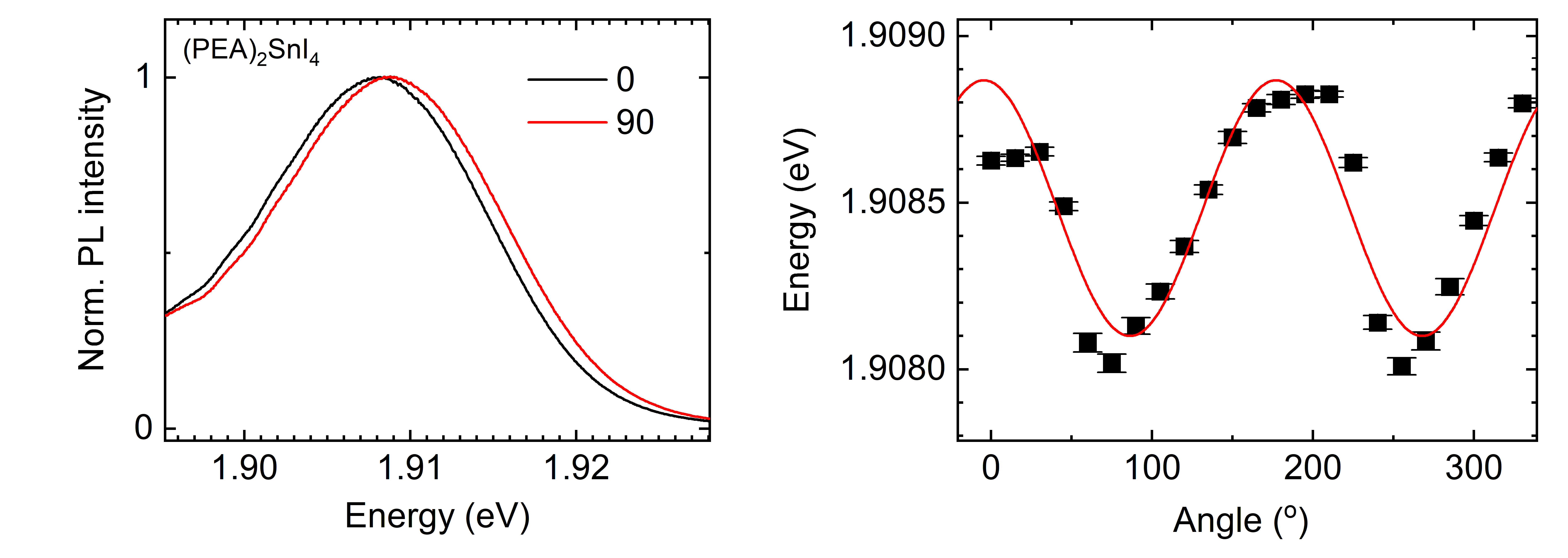}
    \label{fig:SI_peasni_polPL}
\end{figure*}

\begin{figure}
    \centering
    \caption{\textbf{I/E$_b$ ratio in the function of $\beta$ and $\delta$ distortion angles.}\newline
    The ratio of the exchange energy $I$ and exciton binding energy E$_b$ in the function of a) in-plane distortion angle $\beta$ and b) out-of-plane corrugation angle $\delta$. The figure serves as a support to Figure\,\ref{Ofig:fig5_theory} of the main text, where I/E$_b$ is plotted in the function of $\beta+\delta$.  The dashed lines are guides to the eye.
    }
    \includegraphics[width=0.8\linewidth]{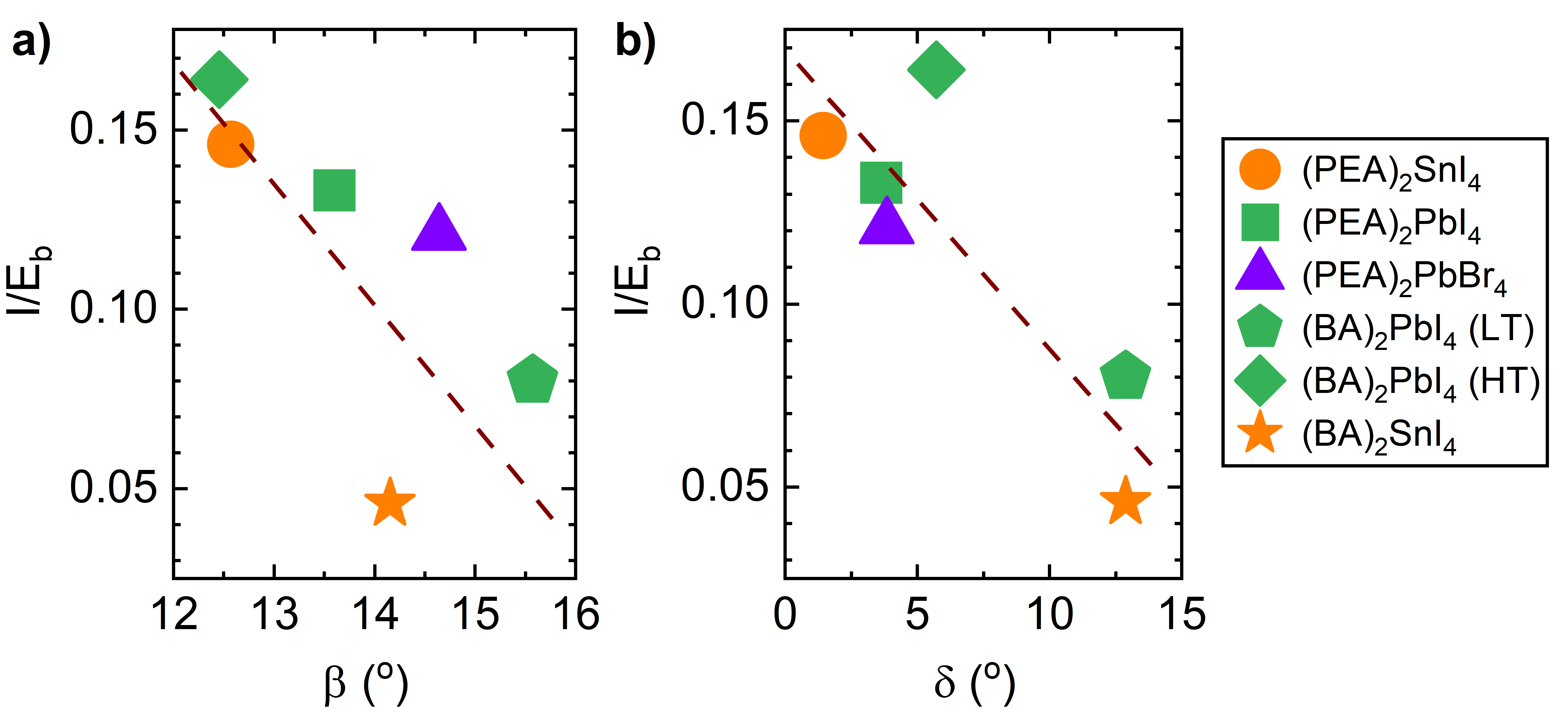}
    \label{fig:SI_individual_exchange}
\end{figure}

\begin{figure}
    \centering
    \caption{\textbf{Impact of corrugation on the bright-bright splitting.}\newline
    The ratio of the energy splitting between the bright in-plane excitons $\Delta_{XY}$ and exciton binding energy E$_b$ in the function of out-of-plane corrugation $\delta$. The open symbol stands for the absolut $\Delta_{XY}$ value for LT (BA)$_2$PbI$_4$. 
    }
    \includegraphics[width=0.6\linewidth]{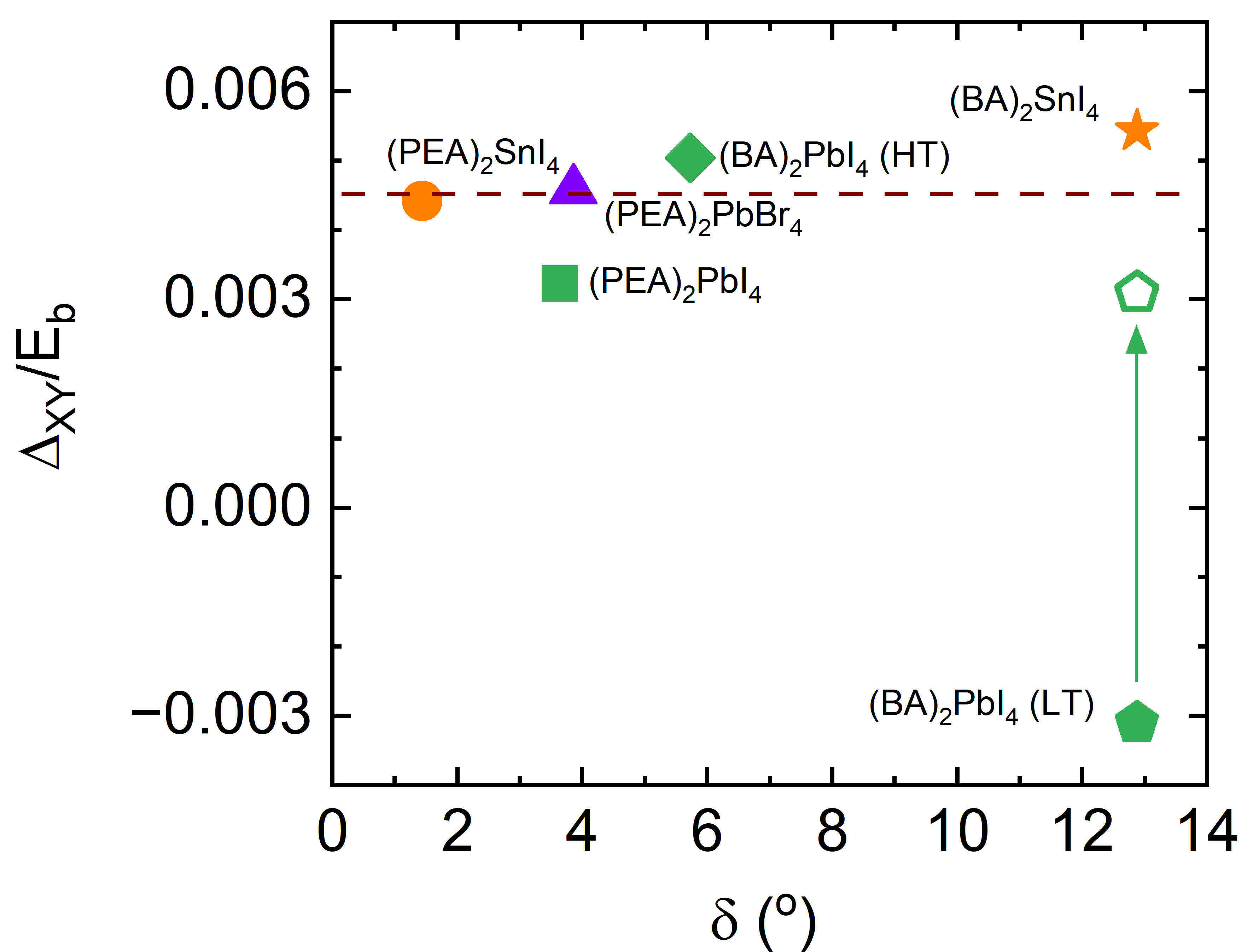}
    \label{fig:SI_Dxy_delta}
\end{figure}

\clearpage
\section{SUPPLEMENTARY TABLES}
\begin{table*}[ht]
\centering
\caption{Summary of $g$-factor measured in Voigt geometry. From the left: effective $g$-factors for longitudinal ($g_L=g_{e\perp}-g_{h\perp}$) and transverse ($g_T=g_{e\perp}+g_{h\perp}$) states, individual electron $g_{e\perp}$ and hole $g_{h\perp}$ $g$-factors both defined for the direction normal to the \textbf{B} vector.
}
\setlength{\tabcolsep}{7pt}
\renewcommand{\arraystretch}{1.2}
\begin{tabular}{l|cccc}
& $g_L$ & $g_T$ & $g_{e\perp}$ & $g_{h\perp}$  \\ 
\hline
(BA)$_2$SnI$_4$ - LT & \num{3.3 +- 0.3} & \num{3.8 +- 0.1} & \num{3.5 +- 0.4} & \num{0.26+-0.32}     
\\
(BA)$_2$PbI$_4$ - LT  & \num{2.7 +- 0.5} & \num{2.9 +- 0.1} & \num{2.8 +- 0.6} & \num{0.10 +- 0.51}          
\\
(BA)$_2$PbI$_4$ - HT & \num{3.3 +- 0.2} & \num{1.8 +- 0.1} & \num{2.6 +- 0.3} & \num{-0.7+-0.3}     
\\
(PEA)$_2$PbI$_4$ & \num{ 3.6 +- 0.3} & \num{2.0 +- 0.2} & \num{2.8 +- 0.4} & \num{-0.79 +- 0.36}                 
\\
(PEA)$_2$PbBr$_4$ &  \num{3.5\pm0.3} & \num{2.8\pm0.2}  & \num{3.1\pm0.4} & \num{-0.35\pm0.36}    
\end{tabular}
\label{tab:SI_gfactor}
\end{table*}

\begin{table*}[ht]
\centering
\caption{\textbf{Effective mass of charge carriers}\newline
The effective mass of charge carriers $\mu$ used to estimate the 1s exciton rms radius in Figure\,\ref{Ofig:fig6_bohr_radius} of the main text. The $\mu$ for (PEA)$_2$PbI$_4$ is measured from the Landau level spectroscopy\cite{dyksik2021tuning}. The $\mu$ for other compounds is approximated from a phenomenological scaling law\cite{Dyksik2022using}. 
}
\setlength{\tabcolsep}{5pt}
\renewcommand{\arraystretch}{1.1}
\begin{tabular}{l|c}
&$\mu (\unit{m_0})$\\ 
\hline
(PEA)$_2$PbI$_4$ & 0.087\cite{dyksik2021tuning} 
\\
(PEA)$_2$SnI$_4$ & 0.072 
\\
(PEA)$_2$PbBr$_4$ & 0.155 
\\
(BA)$_2$SnI$_4$ & 0.083 
\\
(BA)$_2$PbI$_4$ - LT & 0.126 
\\
(BA)$_2$PbI$_4$ - HT & 0.089 
\end{tabular}

\label{tab:SI_masses}
\end{table*}

\clearpage
\section{Microscopic model of exciton fine structure}
\label{sec:model}

The orbital wavefunctions can be be decomposed into their individual contributions, determined by the spin-orbit coupling. 
The latter leads to a mixing of the orbital composition of the lowest conduction band \cite{Becker2018Bright} with momentum $k$, spin $s$ and   orbital $P_i$
\textcolor{black}{\begin{align}
    \ket{s,c,k} =  a_z\sigma_s\ket{P_z,s,k} + a_x\ket{P_x, \overline{s}, k} + i a_y\sigma_s\ket{P_y, \overline{s}, k} ,
\end{align}}
where $\sigma  = \pm 1$ denotes the up/down spin, respectively. Here $a_i$ determines the contribution of the orbital $P_i$ to the valence band. Equal weighting of the $P_x$, $P_y$ and $P_z$ orbitals corresponds to a perovskite with cubic symmetry ($a_x = a_y=a_z = \frac{1}{\sqrt{3}}$). In the orthorhombic phase the weightings differ \cite{steger2022optical}, however the contributing orbitals remain the same. In particular, $a_x = a_y = \dfrac{1}{\sqrt{2}}\cos \theta$ and $a_z  = \sin \theta$ where $\theta$ is a material specific parameter.
There is no mixing of the valence band $S$ orbitals in any phase.
The exchange Hamiltonian can be written as a 4x4 matrix describing each spin state \cite{Thompson2024phononbottleneck}. As such we need to solve the eigenproblem
\begin{align}
      \begin{pmatrix}
     a_z^2I_z & 0 & 0 & -a_z^2I_z \\
     0 & a_x^2I_x + a_y^2I_y & a_x^2I_x - a_y^2I_y  & 0 \\
     0 & a_x^2I_x - a_y^2I_y  & a_x^2I_x + a_y^2I_y  & 0 \\
     -a_z^2I_z & 0 & 0 & a_z^2I_z \\
  \end{pmatrix}\begin{pmatrix}
D^{\uparrow, \uparrow, n}  \\
D^{\uparrow, \downarrow, n} \\
D^{\downarrow, \uparrow, n}\\
D^{\downarrow, \downarrow, n} 
\end{pmatrix}  = E^n \begin{pmatrix}
D^{\uparrow, \uparrow, n}  \\
D^{\uparrow, \downarrow, n}  \\
D^{\downarrow, \uparrow, n}\\
D^{\downarrow, \downarrow, n} ,
\end{pmatrix} 
\end{align}
where $I_{x,y,z}$ are the exchange integrals for the $P_x$, $P_y$ and $P_z$ orbitals respectively.  Here $D^{\uparrow, \uparrow, n} $ describes the spin contribution to each exciton band $n$ with energy $E^n$.
For the tetragonal system $a_x = a_y$ and $I_x = I_y$ and  leading to the energy landscape
\begin{align*}
    E^0 &= 0 \quad\text{(dark)} \\
    E^1&=  2 a_x^2 I_x  \quad\text{(LCP)} \\
    E^2&=  2 a_x^2 I_x  \quad\text{(RCP)}\\
    E^3&=  2 a_z^2 I_z  \quad\text{(Z)}
\end{align*}

In contrast, if $a_x\neq a_y$ or $I_x \neq I_y$ we have 
\begin{align*}
        E^0 &= 0 \quad\text{(dark)} \\
    E^1&=  2 a_x^2 I_x  \quad\text{(X)} \\
    E^2&=  2 a_y^2 I_y  \quad\text{(Y)}\\
    E^3&=  2 a_z^2 I_z  \quad\text{(Z)}
\end{align*}

The bright-dark splitting as defined here, is the splitting between the bright $X$-polarised exciton ($n=1$) and the dark ($n=0$) exciton, as it is these that mix when a magnetic field is applied. We also know by the normalisation of the wavefucntion that $a_x^2+a_y^2+a_z^2 = 1$. Assuming that $I_x = I_y=I_z=I$ and that relative orbital contributions to the band dictate the ordering of states, combined with the normalisation condition allows us to extract $I, a_{x,y,z}$ from the equation from the exciton resonances. In particular we solve the system of equations
\begin{align*}
    I a_x^2 &= \Delta_{BD} \\
    I (a_y^2-a_z^2) &= \Delta_{XY} \\
      I (a_z^2-a_y^2) &= \Delta_{BZ}\\
      a_x^2+a_y^2+a_z^2 &=1,
\end{align*}
in order to uniquely determine $I, a_{x,y,z}$ as discussed in the main text.


The impact of the magnetic field can be included by considering a new 4x4 Hamiltonian whose basis is the exciton states in the absence of a magnetic field \cite{Thompson2024phononbottleneck}
\begin{align}
H_X(B)=\begin{pmatrix}
      E_\text{Dark} & 0 & \dfrac{1}{2}g_L \mu_B B & 0 \\
        0 & E_\text{Z} & 0 & \dfrac{1}{2}g_T \mu_B B\\
     \dfrac{1}{2} g_L \mu_B B& 0 & E_\text{X} & 0 \\
        0& \dfrac{1}{2}g_T \mu_B B & 0 & E_\text{Y} \\
    \end{pmatrix}
\end{align}
where $g_L = g_c-g_v$  and $g_T = g_c+g_v$ depend on the g-factors $g_{c(v)}$ of the exciton-weighted electron (hole) bands \cite{Thompson2024phononbottleneck}.
Solving the matrix above, we obtain the eigenvalues
\begin{align}
    &E_{0L}(B) =  \dfrac{1}{2}\left(E_\text{Dark} +E_\text{X} +\sqrt{(E_\text{Dark} -E_\text{X})^2 + (g_L \mu_B B)^2} \right) \\
       &E_{1L}(B) = \dfrac{1}{2}\left(E_\text{Dark} +E_\text{X} -\sqrt{(E_\text{Dark} -E_\text{X})^2 + (g_L \mu_B B)^2} \right)\\
     &E_{0T}(B) =  \dfrac{1}{2}\left(E_\text{Y} + E_\text{Z} -\sqrt{(E_\text{Y} - E_\text{Z})^2 + (g_T \mu_B B)^2} \right) \\
       &E_{1T}(B) =  \dfrac{1}{2}\left(E_\text{Y} + E_\text{Z} +\sqrt{(E_\text{Y} - E_\text{Z})^2 + (g_T \mu_B B)^2} \right) \\
\end{align}

The corresponding eigenstates are
\begin{align}
    &\psi_{0L}(B) = \dfrac{2}{\sqrt{\mathcal{N}}}\begin{pmatrix}
     \dfrac{ E_{0L}(B)-E_X}{g_L \mu_B B/2}\\
        0\\
        1\\
        0
    \end{pmatrix} \\
    & \psi_{1L}(B) = \dfrac{2}{\sqrt{\mathcal{N}}}\begin{pmatrix}
  \dfrac{ E_{1L}(B)-E_X}{g_L \mu_B B/2}\\
        0\\
        1\\
        0
    \end{pmatrix} \\
        &\psi_{0T}(B) = \dfrac{2}{\sqrt{\mathcal{N}}}\begin{pmatrix}
        0\\
     \dfrac{ E_{0T}(B)-E_Y}{g_T \mu_B B/2}\\
        0\\
        1
    \end{pmatrix} \\
      &\psi_{1T}(B) = \dfrac{2}{\sqrt{\mathcal{N}}}\begin{pmatrix}
        0\\
     \dfrac{ E_{1T}(B)-E_Y}{g_T \mu_B B/2}\\
        0\\
        1
        \end{pmatrix}
\end{align}

\bibliography{Biblio}